\documentclass[aps,prd,preprintnumbers,nofootinbib,superscriptaddress,preprint]{revtex4-2}
\usepackage{graphicx}
\usepackage{amssymb}
\usepackage{amsmath}
\usepackage{xcolor}
\usepackage{booktabs}
\usepackage{dcolumn}
\usepackage{ulem}

\graphicspath{{fig/}}

\newcommand{\beq}{\begin{eqnarray}}
\newcommand{\eeq}{\end{eqnarray}}

\def\fsl#1{\setbox0=\hbox{$#1$}
   \dimen0=\wd0
   \setbox1=\hbox{/} \dimen1=\wd1
   \ifdim\dimen0>\dimen1
      \rlap{\hbox to \dimen0{\hfil/\hfil}}
      #1
   \else
      \rlap{\hbox to \dimen1{\hfil$#1$\hfil}}
      /
   \fi}

\begin{document}

\preprint{UTHEP-822, UTCCS-P-179}

\title{An improvement of model-independent method for meson charge radius calculation}

\date{\today}


\author{Kohei~Sato}
\affiliation{Center for Computational Sciences, University of Tsukuba, \\ Tsukuba, Ibaraki 305-8577, Japan}
\affiliation{Office of the President, Seikei University, \\ Musashino, Tokyo 180-8633, Japan}

\author{Hiromasa~Watanabe}
\affiliation{Department of Physics, and Research and Education Center for Natural Sciences, Keio University, \\
4-1-1 Hiyoshi, Yokohama, Kanagawa 223-8521, Japan}

\author{Takeshi~Yamazaki}
\affiliation{Faculty of Pure and Applied Sciences, University of Tsukuba, \\ Tsukuba, Ibaraki 305-8571, Japan}
\affiliation{Center for Computational Sciences, University of Tsukuba, \\ Tsukuba, Ibaraki 305-8577, Japan}

\begin{abstract}
We propose a variant of the model-independent method for determining meson charge radii from spatial moments of correlation functions on the lattice. 
Traditional determinations based on fits to the momentum transfer squared dependence of form factors are subject to systematic uncertainties arising from the choice of fit ansatz. 
By contrast, model-independent methods based on spatial moments provide a useful framework for determining the slope of the form factor without assuming its functional form. Recently, Feng {\it et al.} proposed a model-independent method, which drastically suppresses the finite-volume effect in the charge radius coming from higher-order contributions of the expansion of the form factor with respect to the momentum transfer squared.
In this work, we introduce an auxiliary function of the momentum transfer squared and reformulate the method in terms of its product with the form factor, rather than the form factor itself, thereby further suppressing higher-order contributions, notably in cases of small volume and large radius. 
In particular, we investigate quadratic and logarithmic forms as practical choices for this auxiliary function. 
Applying this method to mock data based on a monopole form factor, as well as to actual lattice QCD data using $N_f=2+1$ gauge ensembles at $m_\pi \simeq 0.5$ and $0.3$ GeV, we find that it reduces residual finite-volume effects and provides an effective framework for meson charge radius determinations.
\end{abstract}

\maketitle

\section{Introduction}

Accurate determination of hadronic charge radii from first-principles lattice Quantum Chromodynamics (QCD) calculations has become indispensable in modern particle physics.
Notably, discrepancies such as the proton size puzzle~\cite{Pohl:2010zza,ParticleDataGroup:2024cfk} highlight the critical need for precise calculations based on reliable theoretical frameworks.
In this context, lattice QCD, which provides a systematically improvable and nonperturbative framework based on the QCD Lagrangian, plays a pivotal role in validating experimental findings and refining theoretical understanding of hadron structure.

Traditionally, lattice-based extraction of charge radii relies on fitting numerical data for form factors to predetermined analytical models, such as monopole or polynomial functions of the momentum transfer squared~\cite{Boyle:2008yd,Aoki:2015pba,Koponen:2015tkr,ETM:2017wqc,Wang:2020nbf,Gao:2021xsm}.
Although convenient, this approach inherently introduces systematic uncertainties arising from model assumptions.
Indeed, choosing different functional forms can yield different radius results, making accurate determination problematic.
Consequently, developing robust methodologies independent of analytic modeling assumptions is vital to reducing systematic uncertainties.

Model-independent approaches have a long pedigree.
These, which are based on moments in position space, were first introduced in the 1990s to determine the slope of the Isgur-Wise function at zero recoil without assuming any specific ansatz~\cite{Aglietti:1994nx,Lellouch:1994zu}, then extended to study the nucleon form factors~\cite{Alexandrou:2015spa,Bouchard:2016gmc,Alexandrou:2016rbj,Hasan:2017wwt,Alexandrou:2020aja,Ishikawa:2021eut} and meson correlators~\cite{deDivitiis:2012vs}.
It was found, however, that a significant finite-volume effect exists
in the naive method in the previous works~\cite{Lellouch:1994zu,Alexandrou:2016rbj, Ishikawa:2021eut}.

Feng {\it et al.} pointed out that, when the naive method based on moments in position space employed in these studies is applied to the calculation of the pion charge radius, significant finite-volume effects arise~\cite{Feng:2019geu}.
They proposed a method that combines higher-order moments to substantially reduce these effects.
While this approach suppresses finite-volume effects compared to the naive method, systematic biases can arise in certain cases~\cite{Sato:2022qee,Sato:2023vnb}.
We therefore investigate this issue in this paper.

This paper is organized as follows: Section~\ref{sec:Model-independent methods} details both the original and our improved model-independent methods, clarifying their theoretical foundations and practical implementations. Validation tests using mock data are presented in Sec.~\ref{sec:mockup}. Section~\ref{sec:Numerical validation with actual lattice QCD configurations} presents detailed numerical results performed on actual lattice QCD ensembles. Finally, Section~\ref{Summary} summarizes our findings and provides insights into future directions.

We assume that all quantities are written in lattice units, and set the lattice spacing $a = 1$ in the following.

\section{Model-independent methods}
\label{sec:Model-independent methods}

In this section, to explain model-independent methods, we consider the infinite temporal direction for simplicity.

\subsection{Basics of the model-independent method}

First, we consider an infinite volume in one spatial dimension, together with the temporal direction.
A function $F(p^2)$ is assumed to satisfy $F(0) = 1$, and we aim to determine its first derivative.
The correlator $C_\infty(x,t)$ is given by the following Fourier transformation,
\begin{equation}
    C_\infty(x,t) = \int \frac{dp}{2\pi} F(p^2) \Delta(p^2,t) e^{ipx},
    \label{eq:def_corr_infv}
\end{equation}
where $\Delta(p^2,t)$ is a known function with $\Delta(0,t)$ = 1.
The spatial moment of $C_\infty(x,t)$ with $x^2$ can be expressed by the derivative of $F(p^2) \Delta(p^2,t)$ with respect to $p$ as
\begin{eqnarray}
    C_\infty^{\rm mom}(t) &=& \int dx\, x^2 C_\infty(x,t) \\
    &=& \int dx\, x^2  \int \frac{dp}{2\pi} F(p^2) \Delta(p^2,t) e^{ipx}\\
    &=& - \int dx \int \frac{dp}{2\pi} F(p^2) \Delta(p^2,t) 
    \left[ \frac{d^2 e^{ipx}}{dp^2} \right]\\
    &=& - \int dp\,  F(p^2) \Delta(p^2,t) 
    \left[ \frac{d^2 \delta(p)}{dp^2} \right]\\
    &=& - \frac{d^2}{dp^2} \left(F(p^2) \Delta(p^2,t)\right)_{p^2=0} ,
    \label{eq:rel_mnt_der_infv}
\end{eqnarray}
where the integration by parts is used in the last line.
Thus, in the infinite volume, the first derivative of $F(p^2)$ with respect to $p^2$ at $p^2 = 0$ is obtained from the moment $C_\infty^{\rm mom}(t)$ and the derivative of the known function $\Delta(p^2,t)$,
\begin{equation}
    \left.\frac{d F(p^2)}{dp^2}\right|_{p^2=0} = 
    - \frac{1}{2} C_\infty^{\rm mom}(t) - 
    \left.\frac{d \Delta(p^2,t)}{dp^2}\right|_{p^2=0} .
    \label{eq:mod-indep_infv}
\end{equation}

Next, the relation between the moment and the derivative in Eq.~(\ref{eq:rel_mnt_der_infv}) is applied to calculation on a lattice in one spatial dimension with periodic boundary conditions.
The spatial length is given by $L$.
On a lattice, the integral in Eq.~(\ref{eq:def_corr_infv}) is replaced by a summation of the momentum $p$ as,
\begin{equation}
    C(x,t) = \frac{1}{L}\sum_{p} F(p^2) \Delta(p,t) e^{-ipx},
\end{equation}
where $p = (2 \pi/L) \cdot n_p$ with $-L/2+1 \le n_p \le L/2$ and $n_p$ being an integer.
Using the correlator, the $2k$th-order spatial moment on a lattice, $C^{(k)}(t)$, is written as
\begin{equation}
    C^{(k)}(t) = \sum_x x^{2k} C(x,t) = 
    \sum_{p} F(p^2) \Delta(p^2,t) T_k(p),
    \label{eq:def_mnt_fv}
\end{equation}
where
\begin{equation}
    T_k(p) = \frac{1}{L}\sum_x x^{2k} e^{-ipx}
\end{equation}
with discrete values of $x$ satisfying $-L/2 + 1 \le x \le L/2$.
Similar to the infinite-volume case in Eq.~(\ref{eq:mod-indep_infv}), the derivative of $F(p^2)$ can be evaluated from $C^{(1)}(t)$ and the derivative of the known function $\Delta(p^2,t)$ as,
\begin{equation}
    \left.\frac{d F(p^2)}{dp^2}\right|_{p^2=0} = 
    - \frac{1}{2} D^{(1)}(t) - 
    \left.\frac{d \Delta(p^2,t)}{dp^2}\right|_{p^2=0} ,
    \label{eq:naive_mod-indep_fv}
\end{equation}
where 
\begin{equation}
    D^{(k)}(t) = \frac{C^{(k)}(t)}{C^{(0)}(t)} .
    \label{eq:naive_def_dt}
\end{equation}
The normalization factor in Eq.~(\ref{eq:naive_def_dt}) is required to satisfy $D^{(0)}(t) = 1$ at each $t$.
This method is called a model-independent method, because the derivative of $F(p^2)$ is calculated without a model assumption for $p^2$ dependence of $F(p^2)$.
However, it was found in Refs.~\cite{Lellouch:1994zu,Alexandrou:2016rbj} that a large finite-volume effect exists in this method.

\subsection{Linear combination method by Feng {\it et al.}}
\label{sec:improved_method}

Feng {\it et al.}~\cite{Feng:2019geu} proposed an improved method for the naive model-independent method in Eq.~(\ref{eq:naive_mod-indep_fv}).
Although their method was described using a correlator in three-dimensional space in Ref.~\cite{Feng:2019geu}, we discuss the case of one spatial dimension as follows for simplicity.

In their method, the spatial moment $D^{(k)}(t)$ is represented by the Taylor series of $F(p^2)$ given by
\begin{equation}
    F(p^2) = \sum_{n\ge0} f_n p^{2n}
    \label{eq:Fp_expansion}
\end{equation}
with $f_0 = F(0) = 1$ as,
\begin{eqnarray}
    D^{(k)}(t) &=& \frac{1}{C^{(0)}(t)}\sum_x x^{2k} C(x,t)\\ 
    &=& 
    \sum_{p} \sum_n f_n p^{2n} \Delta(p^2,t) T_k(p)\\
    &=& 
    \sum_n f_n \overline{\beta}_{nk}(t),
    \label{eq:imp_method_expansion}
\end{eqnarray}
where
\begin{equation}
    \overline{\beta}_{nk}(t) = \sum_p p^{2n} \Delta(p^2,t) T_k(p) .
    \label{eq:def_beta-bar}
\end{equation}
It should be noted that $\overline{\beta}_{nk}(t)$ is written in terms of only known functions, {\it i.e.,} it does not contain the expansion coefficients of $F(p^2)$.
Then, the first derivative of $F(p^2)$ is evaluated from $\overline{R}(t)$, which is a linear combination of $C^{(1)}(t)$ and $C^{(2)}(t)$ with parameters $\overline{\alpha}_0,\overline{\alpha}_1,\overline{\alpha}_2$ given by
\begin{equation}
    \overline{R}(t) = \overline{\alpha}_0 + \overline{\alpha}_1 D^{(1)}(t) + 
    \overline{\alpha}_2 D^{(2)}(t) .
    \label{eq:imp_method_def_R}
\end{equation}
The parameters are determined by the condition that the contribution proportional to $f_2$ is removed, as
\begin{eqnarray}
    \overline{\alpha}_0 + \overline{\alpha}_1 \overline{\beta}_{01}(t) + \overline{\alpha}_2 \overline{\beta}_{02}(t) &=& 0 ,
    \label{eq:condition_1}\\
    \overline{\alpha}_1 \overline{\beta}_{11}(t) + \overline{\alpha}_2 \overline{\beta}_{12}(t) &=& 1 ,
    \label{eq:condition_2}\\
    \overline{\alpha}_1 \overline{\beta}_{21}(t) + \overline{\alpha}_2 \overline{\beta}_{22}(t) &=& 0 .
    \label{eq:condition_3}
\end{eqnarray}
This is nothing but solving the linear equations given by
\begin{equation}
    \left(
    \begin{array}{cc}
        \overline{\beta}_{11}(t) & \overline{\beta}_{21}(t) \\
        \overline{\beta}_{12}(t) & \overline{\beta}_{22}(t) 
    \end{array}
    \right)
    \left(
    \begin{array}{c}
        f_1\\
        f_2
    \end{array}
    \right) =
    \left(
    \begin{array}{c}
        D^{(1)}(t) - \overline{\beta}_{01}(t)\\
        D^{(2)}(t) - \overline{\beta}_{02}(t)
    \end{array}
    \right) 
\end{equation}
with respect to $f_1$.
Under the conditions in Eqs.~(\ref{eq:condition_1})--(\ref{eq:condition_3}), $\overline{R}(t)$ is given by $f_1$ and higher-order contributions for $n\ge3$, which are the source of the finite-volume effect, as,
\begin{equation}
    \overline{R}(t) = f_1 + \sum_{n\ge3} f_n \left(
    \overline{\alpha}_1\overline{\beta}_{n1}(t)+\overline{\alpha}_2\overline{\beta}_{n2}(t)\right) .
    \label{eq:imp_method_R}
\end{equation}

This method reduces the finite-volume effect significantly compared to the naive model-independent method described in Eq.~(\ref{eq:naive_mod-indep_fv}).
The reason is that the finite-volume effect stemming from the derivative of $\Delta(p^2,t)$ and proportional to $f_2$ completely disappears.
Using this method, Feng {\it et al.} successfully reproduced the experimental value of the pion charge radius in calculations at the physical point~\cite{Feng:2019geu}.
We shall call this method the linear combination method in the following.

\subsection{Our improved model-independent method}
\label{sec:our_method}

In cases with large $|f_1|$ on a small volume in one-dimensional space, the finite-volume effect can be problematic even with the linear combination method; an example is discussed in the next section.
To avoid the finite-volume effect, we propose a new improved model-independent method.
The strategy of our method is to reduce the higher-order contribution in Eq.~(\ref{eq:imp_method_R}) by improving the convergence of the Taylor expansion in $\overline{R}(t)$.
In other words, we aim to decrease the absolute values of the higher-order coefficients of the Taylor expansion by replacing the expanded function with a more appropriate one.
For this purpose, we define $S(p^2)$ and expand it as
\begin{equation}
    S(p^2) = F(p^2)G(p^2) = \sum_{n=0} s_n p^{2n}
    \label{eq:our_imp_method_expansion}
\end{equation}
with $s_0 = F(0) G(0) = 1$ by introducing a function $G(p^2)$ with $G(0) = 1$.
Using $1 = G(p^2)/G(p^2)$ in Eq.~(\ref{eq:def_mnt_fv}) and the expansion of Eq.~(\ref{eq:our_imp_method_expansion}), one obtains
\begin{eqnarray}
    D^{(k)}(t) &=& \sum_p S(p^2)\Delta(p^2,t)T_k(p)/G(p^2)
    \label{eq:our_method_Dt_expand_1}\\
    &=&
    \sum_n s_n \beta_{nk}(t) ,
    \label{eq:our_method_Dt_expand_2}
\end{eqnarray}
where
\begin{equation}
    \beta_{nk}(t) = \sum_p p^{2n} \Delta(p^2,t) T_k(p) / G(p^2) .
    \label{eq:our_method_def_beta}
\end{equation}
Similar to $\overline{\beta}_{nk}(t)$ in Eq.~(\ref{eq:def_beta-bar}), it is given by only the known functions.
Following the procedure in the linear combination method, the linear combination $R(t)$ is defined as
\begin{equation}
    R(t) = \alpha_0 + \alpha_1 D^{(1)}(t) + \alpha_2 D^{(2)}(t) - g_1,
    \label{eq:our_method_def_Rt}
\end{equation}
where $g_1$ is evaluated from the Taylor expansion of $G(p^2)$,
\begin{equation}
    G(p^2) = 1 + \sum_{n\ge1}g_n p^{2n} 
\end{equation}
and the parameters $\alpha_i$ satisfy
\begin{eqnarray}
    \alpha_0 + \alpha_1 \beta_{01}(t) + \alpha_2 \beta_{02}(t) &=& 0 ,
    \label{eq:our_method_condition_1}
    \\
    \alpha_1 \beta_{11}(t) + \alpha_2 \beta_{12}(t) &=& 1 ,
    \label{eq:our_method_condition_2}
    \\
    \alpha_1 \beta_{21}(t) + \alpha_2 \beta_{22}(t) &=& 0 .
    \label{eq:our_method_condition_3}
\end{eqnarray}
In this case, $R(t)$ is written in terms of the first derivative of $F(p^2)$, $f_1 = s_1 - g_1$, and the modified higher-order contribution relative to that in Eq.~(\ref{eq:imp_method_R}),
\begin{equation}
    R(t) = f_1 + \sum_{n\ge3}s_n\left(\alpha_1\beta_{n1}(t)+\alpha_2\beta_{n2}(t)\right) .
    \label{eq:our_method_R}
\end{equation}

It is expected that the higher-order contribution in Eq.~(\ref{eq:our_method_R}) is smaller than that of the linear combination method in Eq.~(\ref{eq:imp_method_R}), if $|s_n| < |f_n|$ for $n\ge 3$.
In order to satisfy this requirement, one needs to choose a proper function $G(p^2)$ for each calculation.

One example for the choice of $G(p^2)$ is the exact case, $G(p^2) = 1/F(p^2)$.
In this case, the higher-order contribution as well as the $s_1$ term do not exist; in other words, $s_n = 0$ for $n\ge 1$, and then one can easily calculate $f_1$ as $f_1 = -g_1$.
As expected from the exact case, if one chooses $G(p^2) \approx 1/F(p^2)$ in a small $p^2$ region, the higher-order contribution becomes smaller than that in the linear combination method.
We will demonstrate a reduction of the higher-order contribution by using mock data in the next section.

Of course, in this method, we need to estimate a systematic error arising from the choice of $G(p^2)$, {\it e.g.}, by varying parameters in $G(p^2)$.
An example of this estimate will be discussed in a later section.

\section{Results with mock data}
\label{sec:mockup}

In this section, we discuss the results obtained from the linear combination method and our method explained above in several cases using mock data in two-dimensional spacetime, with one spatial and one temporal dimension.

Since we focus on the calculation of the meson charge radius, especially for the pion case, the mock data are generated with a monopole form factor given by
\begin{equation}
    F(Q^2) = \frac{1}{1 + A Q^2},
    \label{eq:mockup_form_factor}
\end{equation}
where $A$ is an input parameter and the momentum transfer $Q^2 = 2 m_\pi (E_\pi(p) - m_\pi)$ with $E_\pi(p) = \sqrt{p^2+m_\pi^2}$, $p$ and $m_\pi$ being a lattice momentum, $p = (2\pi/L)\cdot n_p$, and the pion mass, respectively.
In this test analysis with the mock data, we fix $m_\pi = 0.25$.

The Taylor expansion of $F(Q^2)$ at $Q^2 = 0$ is given by
\begin{equation}
    F(Q^2) = \sum_{n\ge0} \left( -A \right)^n Q^{2n} 
    = \sum_{n\ge0} f_n Q^{2n} .
    \label{eq:monopole_expand}
\end{equation}
We note that the expansion variable is changed from $p^2$ in Eq.~(\ref{eq:Fp_expansion}) to $Q^2$ hereafter.
The mean square charge radius is defined by $\langle r^2\rangle = -6f_1 = 6A$, and with the input value $A$, we will compare $-\overline{R}(t)$ or $-R(t)$ obtained from Eqs.~(\ref{eq:imp_method_def_R}) and (\ref{eq:our_method_def_Rt}), respectively.

Similar to Eq.~(\ref{eq:def_mnt_fv}), the mock data of the $2k$th-order moment of the correlator are calculated by
\begin{eqnarray}
    C^{(k)}(t) &=& \sum_{p} F(Q^2) \Delta(p^2,t) T_k(p) ,
\end{eqnarray}
where
\begin{eqnarray}
    T_k(p) &=& \frac{1}{L} \sum_{x} x^{2k} \cos(p x) ,
    \label{eq:def_T_k}
\end{eqnarray}
and
\begin{eqnarray}
    \Delta(p^2,t) &=& \frac{E_\pi(p)+m_\pi}{2E_\pi(p)}e^{-(E_\pi(p)-m_\pi)t}
    \label{eq:def_Delta_p_t}
\end{eqnarray}
are employed in this test to mimic a realistic calculation of the meson charge radius.

\subsection{Linear combination method}
\label{sec:mockup_improved_method}

As described in Sec.~\ref{sec:improved_method}, the normalized correlator $D^{(k)}(t) = C^{(k)}(t)/C^{(0)}(t)$ is expanded by the Taylor series of $F(Q^2)$ in the linear combination method as in Eq.~(\ref{eq:imp_method_expansion}) with
\begin{equation}
    \overline{\beta}_{nk}(t) = \sum_p Q^{2n}\Delta(p^2,t)T_k(p) .
    \label{eq:linear_method_def_beta-bar_q2}
\end{equation}
The value of $f_1$ can be obtained from the linear combination $\overline{R}(t)$ in Eq.~(\ref{eq:imp_method_def_R}) with the data $D^{(1)}(t)$ and $D^{(2)}(t)$, and a set of parameters that satisfy the conditions in Eqs.~(\ref{eq:condition_1})--(\ref{eq:condition_3}).

\begin{figure}[ht!]
    \includegraphics*[scale=0.55]{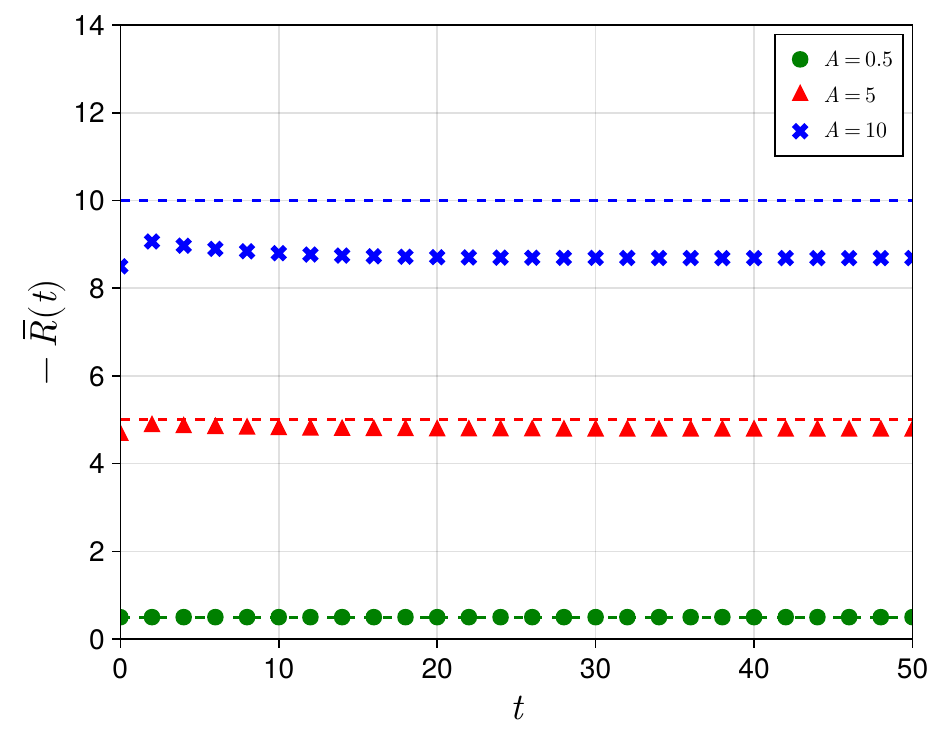}
    \caption{Results of $-\overline{R}(t)$ from the mock data using the linear combination method for $A = 0.5, 5,$ and 10 on $L=32$ as a function of $t$.
    The different symbols correspond to the different $A$ results.
    The horizontal lines denote the input values of $A$.}
    \label{fig:mock_imp_method_M_dep}
\end{figure}

Figure~\ref{fig:mock_imp_method_M_dep} shows the $A$ dependence of the result for $-\overline{R}(t)$ on a fixed volume of $L=32$ as a function of $t$, where $A = 0.5, 5,$ and $10$ are employed.
In the $A=0.5$ case corresponding to a smaller $\langle r^2\rangle$ than those for $A=5$ and $10$, the result for $-\overline{R}(t)$ reasonably agrees with the input value of $A$, which is expressed by a blue horizontal line in the figure.
On the other hand, the deviation of $-\overline{R}(t)$ from the horizontal line appears for the large $A$ cases.
The deviation increases with the value of $A$.
This is because the coefficient of the higher-order contribution in Eq.~(\ref{eq:imp_method_R}) is given by $f_n=(-A)^{n}$ for $n\ge 3$, which worsens the convergence of the expansion of $-\overline{R}(t)$ as $A$ increases on a fixed volume.
In such a case, the higher-order contributions are non-negligible, and hence cause a systematic error in the linear combination method.

In a later section, we will analyze actual lattice QCD data, whose parameters, such as $m_\pi$, $A$, and $L$, are similar to those in the case with $A=5$ shown in the figure.
In this case, $-\overline{R}(t)$ in a large-$t$ region is roughly 5\% smaller than the input $A$.
We will compare the size of the discrepancy of $-\overline{R}(t)$ with that in the actual lattice QCD data.

\begin{figure}[ht!]
    \includegraphics*[scale=0.55]{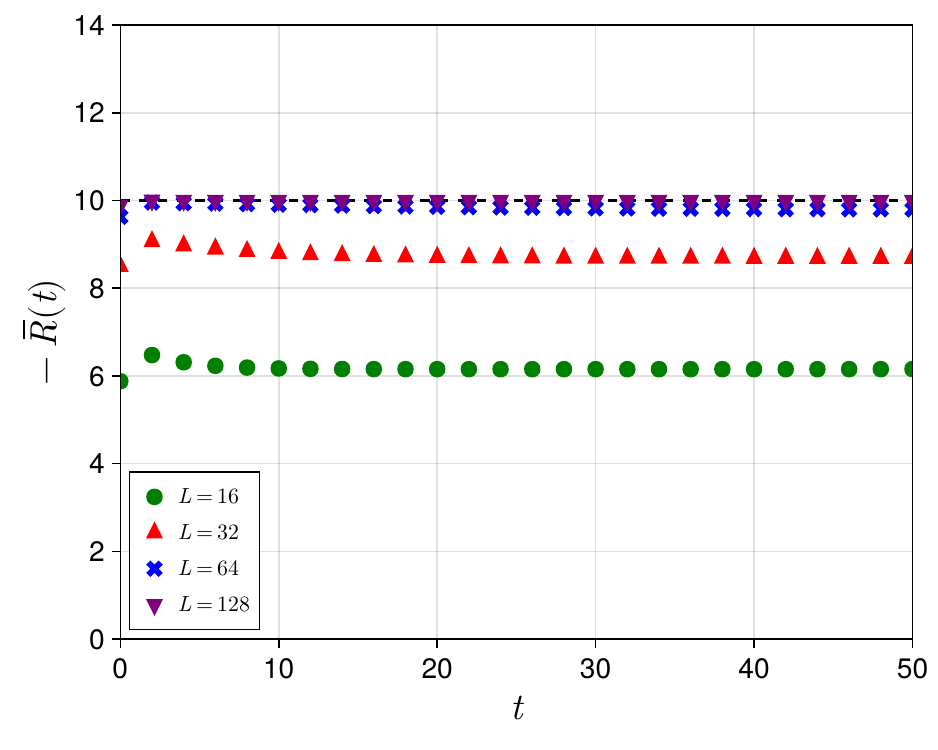}
    \caption{Results of $-\overline{R}(t)$ from the mock data using the linear combination method for $L = 16, 32, 64,$ and 128 with $A = 10$ as a function of $t$.
    The different symbols correspond to the different $L$ results.
    The horizontal line denotes the input value of $A$.}
    \label{fig:mock_imp_method_L_dep}
\end{figure}

An example of the $L$ dependence of $-\overline{R}(t)$ is presented in Fig.~\ref{fig:mock_imp_method_L_dep} with $A=10$, which is interpreted as a large $\langle r^2\rangle$ case.
On a smaller volume, the deviation of $-\overline{R}(t)$ from the input value of $A$ is large, while it decreases as the volume increases.
This deviation is caused by a sizable finite-volume effect in $\overline{\beta}_{nk}(t)$ for $n\ge 3$ and $k=1,2$ in Eq.~(\ref{eq:imp_method_R}).
To reduce the finite-volume effect in large $A$ cases, a larger volume is necessary.

From these mock-data analyses, it is concluded that the finite-volume effect arising from the higher-order contribution in Eq.~(\ref{eq:imp_method_R}) is not negligible for large $A$ and/or small $L$ ($A\gtrsim 5$ or $L\lesssim 64$ in our case), even if we employ the linear combination method.
It is also found that the finite-volume effect in the linear combination method always causes an underestimation of $-f_1$ in the monopole form case.

In Appendix~\ref{app:sys_err_mockup}, we estimate the size of the finite-volume effect in model-independent methods using several approaches: the naive method, the linear combination method, and our two methods explained below.
It should be noted that the finite-volume effect becomes significantly larger when the same data are analyzed by the naive method.
A comparison of the different methods, as well as the $m_\pi$ dependence of each method, is also discussed in the appendix.
In the same appendix, using the mock data, we show that the linear combination method works well at the physical $m_\pi$ parameter with the spatial extent of 6 fm, where the higher-order contributions are well suppressed to below 1\%.

\subsection{Our method}
\label{sec:method:our_method}

As explained in Sec.~\ref{sec:our_method}, we introduce a proper function $G(Q^2)$ in our method.
Using $G(Q^2)$, the $2k$th-order moment correlator is expanded as 
\begin{eqnarray}
    D^{(k)}(t) &=& \sum_p S(Q^2)\Delta(p^2,t)T_k(p)/G(Q^2)\\
    &=& \sum_{n\ge0} s_n \beta_{nk}(t),
\end{eqnarray}
where
\begin{eqnarray}
    S(Q^2) &=& F(Q^2)G(Q^2) = \sum_{n\ge0} s_n Q^{2n} , \\
    \beta_{nk}(t) &=& \sum_p Q^{2n} \Delta(p^2,t)T_k(p)/G(Q^2) ,
    \label{eq:our_method_def_beta_q2}
\end{eqnarray}
as in Eqs.~(\ref{eq:our_method_Dt_expand_1})--(\ref{eq:our_method_def_beta}).
We assume $s_0 = 1$, {\it i.e.}, $G(0) = 1$.
As shown in Eqs.~(\ref{eq:our_method_def_Rt}) and (\ref{eq:our_method_R}), to obtain $-f_1$, the linear combination $-R(t)$ is constructed from $D^{(k)}(t)$ for $k=1,2$ and the parameters $\alpha_i$, with the conditions in Eqs.~(\ref{eq:our_method_condition_1})--(\ref{eq:our_method_condition_3}) and $g_1$ being one of the Taylor expansion coefficients of $G(Q^2)$ defined by
\begin{equation}
    G(Q^2) = 1 + \sum_{n\ge1} g_n Q^{2n}.
\end{equation}
To achieve better convergence in the expansion of $S(Q^2)$ than in that of $F(Q^2)$, {\it i.e.}, improved convergence of the higher-order contributions in Eq.~(\ref{eq:our_method_R}), $G(Q^2)$ should be chosen to reduce $|s_n|$ for $n \ge 3$.

We discuss two examples of $G(Q^2)$ below, which are suitable for meson charge radius calculations.

\subsubsection{Quadratic function}

For this mock data expressed by the monopole form $F(Q^2)$, we choose $G(Q^2)$ to be a quadratic function of $Q^2$:
\begin{equation}
    G(Q^2) = 1 + g_1 Q^2 + g_2 Q^4 ,
\end{equation}
where $g_1$ and $g_2$ are parameters. 
If the parameters satisfy the condition
\begin{equation}
    s_2 = f_2 + f_1 g_1 + g_2 = 0 ,
    \label{eq:our_method_quad_condition_s2}
\end{equation}
the higher-order contribution completely vanishes.
The reason is that 
\begin{equation}
    s_n = f_1 s_{n-1} = 0 ,
\end{equation}
for $n \ge 3$\footnote{In the general case with a monopole form $F(Q^2)$, $s_n = f_1 s_{n-1} + g_n$ for $n>1$.}, where we use $f_n = f_1 f_{n-1}$, which is a special property of the expansion coefficients of the monopole form shown in Eq.~(\ref{eq:monopole_expand}).
This choice eliminates $|s_n|$ for $n \ge 2$, thereby reducing higher-order contamination in Eq.~(\ref{eq:our_method_R}) and enabling a more precise determination of the first-derivative coefficient.

In an actual numerical analysis, the exact values of $f_1$ and $f_2$ in Eq.~(\ref{eq:our_method_quad_condition_s2}) cannot be known a priori, so that the parameters in $G(Q^2)$ need to be determined by a quantity calculated from the data $D^{(k)}(t)$.
A proper quantity is
\begin{equation}
    R_2(t) = \kappa_0 + \kappa_1 D^{(1)}(t) + \kappa_2 D^{(2)}(t),
    \label{eq:def_R2t}
\end{equation}
which corresponds to $s_2$ with a higher-order contribution as,
\begin{equation}
    R_2(t) = s_2 + \sum_{n\ge3} s_n (\kappa_1 \beta_{n1}(t)+\kappa_2\beta_{n2}(t)) .
\end{equation}
The linear combination is similar to $R(t)$ in Eq.~(\ref{eq:our_method_def_Rt}), while it employs a different parameter choice given by
\begin{eqnarray}
    \kappa_0 + \kappa_1 \beta_{01}(t) + \kappa_2 \beta_{02}(t) &=& 0 ,
    \label{eq:our_method_s2_condition_1}
    \\
    \kappa_1 \beta_{11}(t) + \kappa_2 \beta_{12}(t) &=& 0 ,
    \label{eq:our_method_s2_condition_2}
    \\
    \kappa_1 \beta_{21}(t) + \kappa_2 \beta_{22}(t) &=& 1 .
    \label{eq:our_method_s2_condition_3}
\end{eqnarray}
We will demonstrate analyses with the parameters determined from
$R_2(t) = 0$ below.

\begin{figure}[ht!]
    \includegraphics*[scale=0.55]{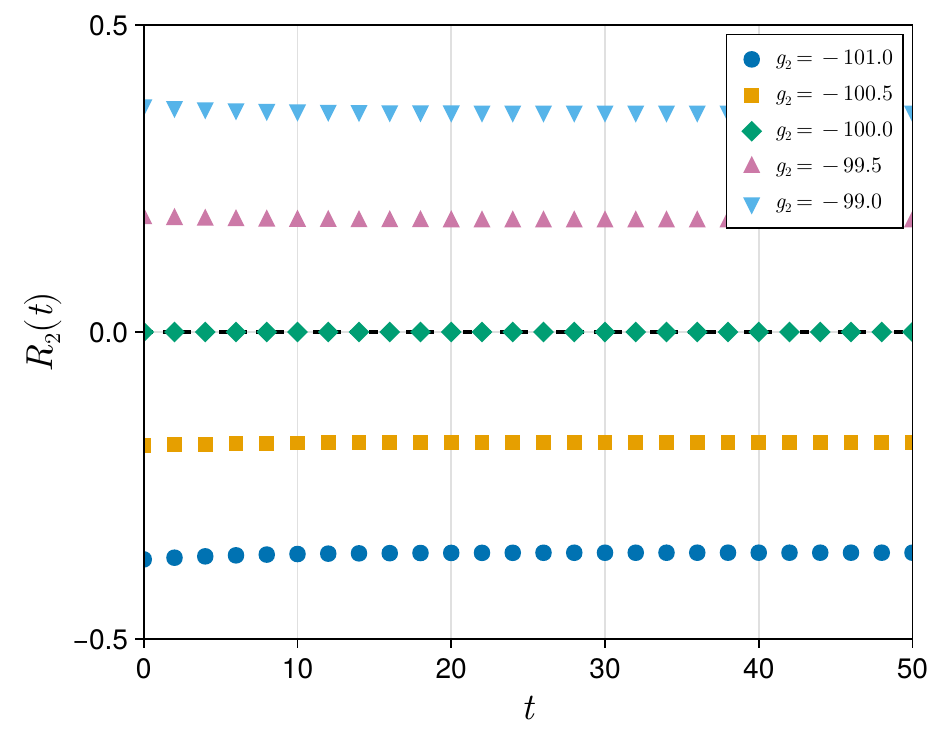}
    \caption{Data of $R_2(t)$ for several values of $g_2$ with fixed $g_1 = 0$ using the quadratic function.}
    \label{fig:mock_our_method_1_quad}
\end{figure}

For a typical example of the analysis with a quadratic function, we adopt $A=10$ and $L=32$ for the mock data $D^{(k)}(t)$.
For the parameter search, we investigate the $g_2$ dependence of $R_2(t)$ with a fixed $g_1$.
Figure~\ref{fig:mock_our_method_1_quad} presents the data of $R_2(t)$
with $g_1 = 0$ and several values of $g_2$, which give smaller values of $R_2(t)$.
Those data behave as a constant with respect to $t$.
Thus, in the following, a value of $R_2(t)$ at $t = t_r=25$ is chosen as a reference to monitor its $g_2$ dependence with a fixed $g_1$.

\begin{figure}[ht!]
    \includegraphics*[scale=0.55]{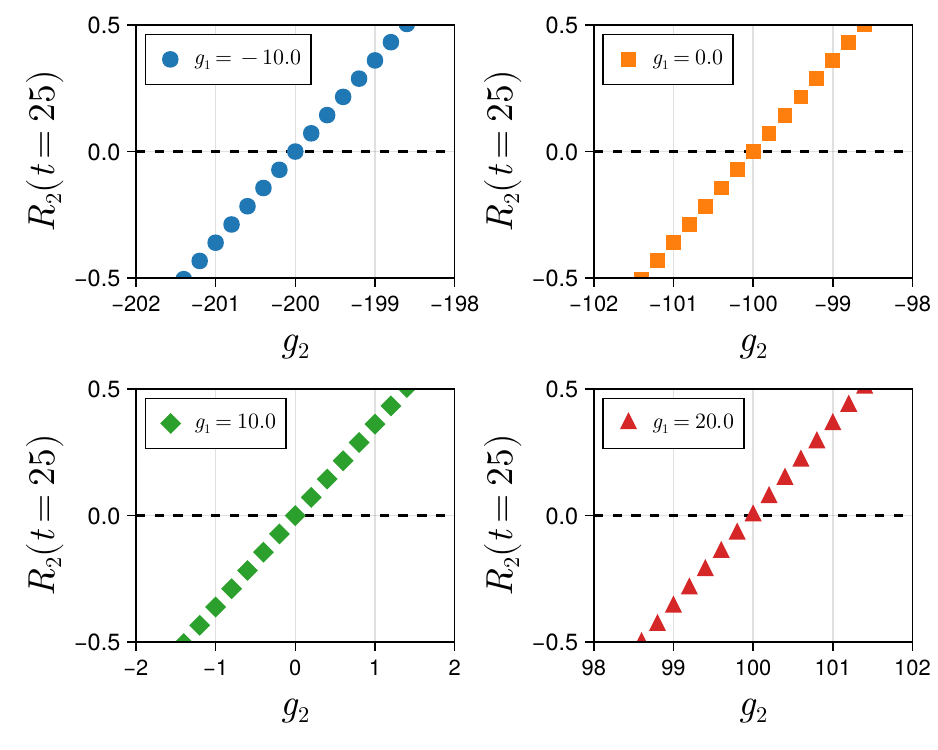}
    \caption{Dependence of $g_2$ near $R_2(t_r) = 0$ with several values of $g_1$ using the quadratic function.}
    \label{fig:mock_our_method_2_quad}
\end{figure}

Figure~\ref{fig:mock_our_method_2_quad} shows that the values of $R_2(t_r)$ in four $g_1$ cases reasonably behave as linear functions of $g_2$ near $R_2(t_r) = 0$.
From this behavior, the value of $g_2$ can be easily determined at $R_2(t_r) = 0$ for each $g_1$.

\begin{figure}[ht!]
\includegraphics*[scale=0.55]{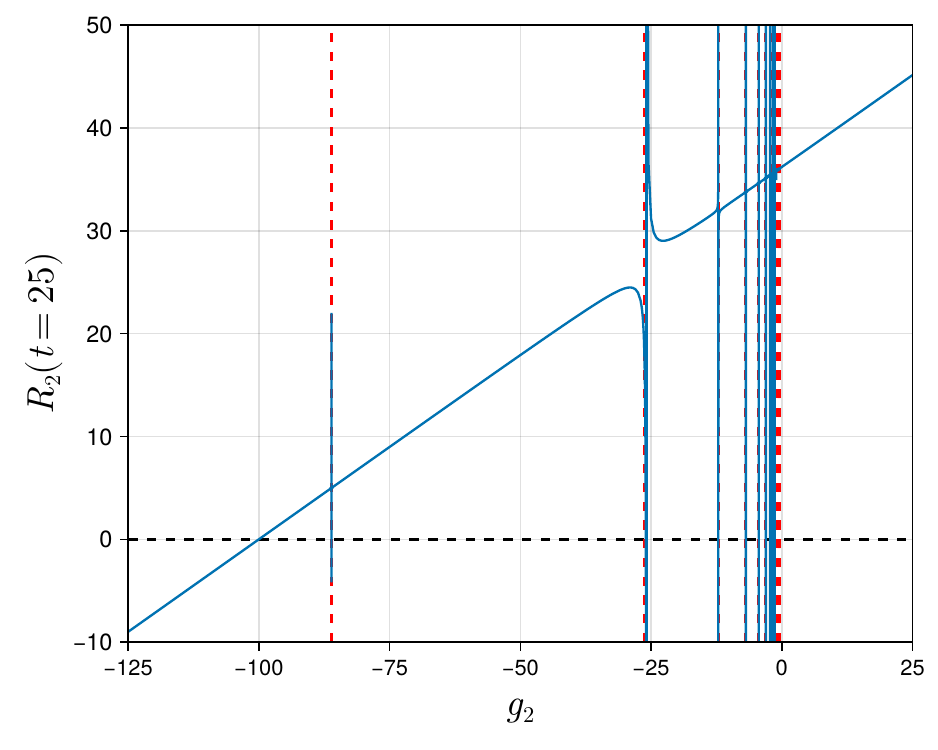}
    \caption{The dependence of $g_2$ on $R_2(t_r)$ with $g_1 = 0$.
    The red dashed line indicates the value of $g_2$ for which $G(Q^2) = 0$, {\it i.e.}, $g_2 = -1/Q^4$.
    One can see that the behavior of the numerical calculation becomes unstable in the vicinity of the red dashed line.
    }
    \label{fig:mock_our_method_3_quad}
\end{figure}

Although we do not encounter a practical problem in determining $g_2$ in actual analyses, we should note that $R_2(t)$ diverges at specific parameter sets for $g_1$ and $g_2$ with fixed $A$, $L$ and $m_\pi$, when $R_2(t)$ is evaluated over a wide range of $g_2$. This is caused by $G(Q^2) = 0$ at some momentum $p^2$ in the denominator of $\beta_{nm}(t)$ in Eq.~(\ref{eq:our_method_def_beta}).
An example is presented in Fig.~\ref{fig:mock_our_method_3_quad} with $g_1 = 0$.
At the values of $g_2$ expressed by the dashed lines, $R_2(t_r)$ diverges due to $G(Q^2) = 0$.
The values of $R_2(t)$ near the divergent points can be zero, although those values are different from what we search for.
In the figure, the correct value of $g_2$ at $R_2(t)=0$, $g_2 = -100$, can be distinguished from the ones near $G(Q^2) = 0$ by investigating  the $g_2$ dependence of $R_2(t)$ and avoiding values of $g_2$ associated with $G(Q^2) = 0$.

\begin{figure}[ht!]
    \includegraphics*[scale=0.55]{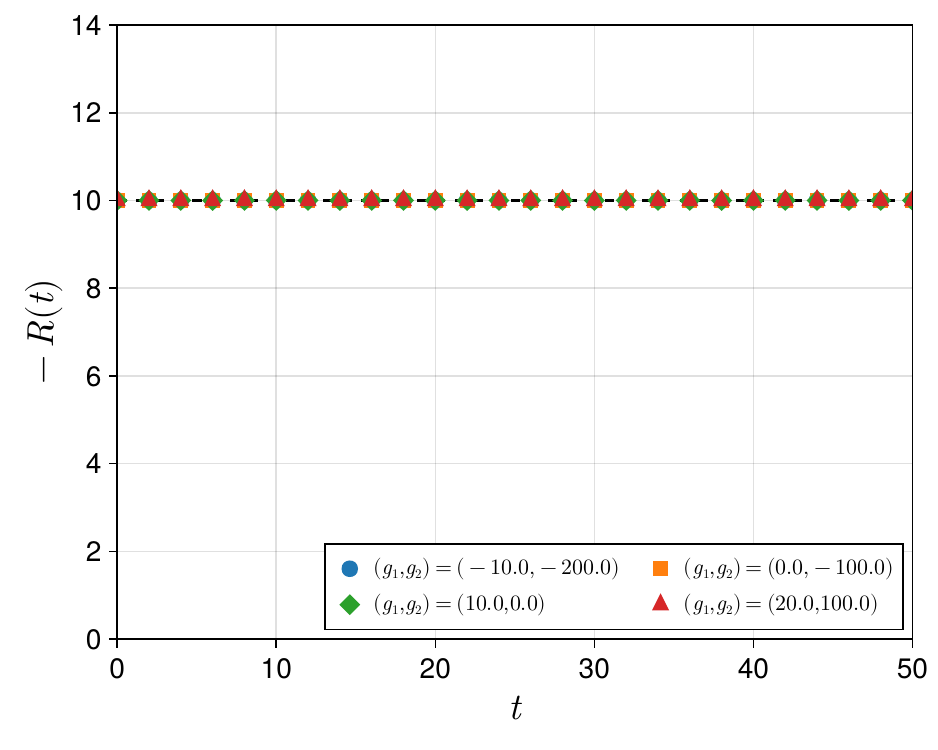}
    \caption{Results of $-R(t)$ with several parameter sets using the quadratic function.
    The dashed line represents the input $A = 10$.}
    \label{fig:mock_our_method_4_quad}
\end{figure}

Using $g_2$ determined from $R_2(t_r)=0$, the value of $R(t)$ is calculated in each $g_1$ case, as presented in Fig.~\ref{fig:mock_our_method_4_quad}.
The results of $-R(t)$ in all the cases reproduce the input value of $A=10$ represented by the dashed line.

\begin{figure}[ht!]
    \includegraphics*[scale=0.55]{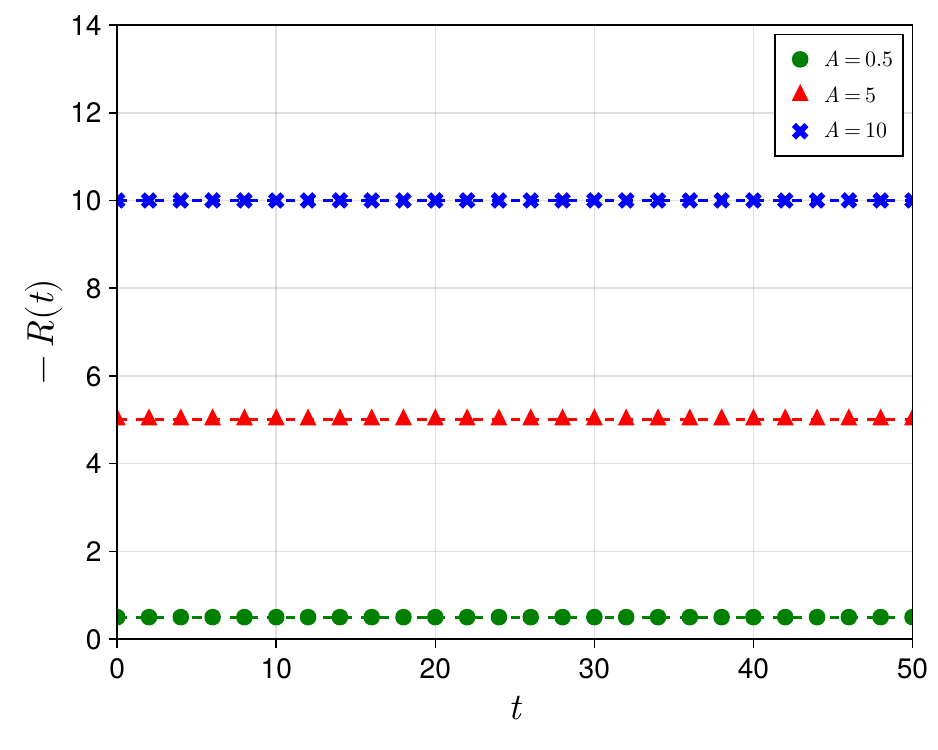}
    \caption{Results of $-R(t)$ from the mock data using our method with the quadratic function for $G(Q^2)$ with $g_1 = 0$ for $A = 0.5, 5,$ and 10 on $L=32$ as a function of $t$.}
    \label{fig:mock_our_method_M_dep_quad}
\end{figure}

\begin{figure}[ht!]
    \includegraphics*[scale=0.55]{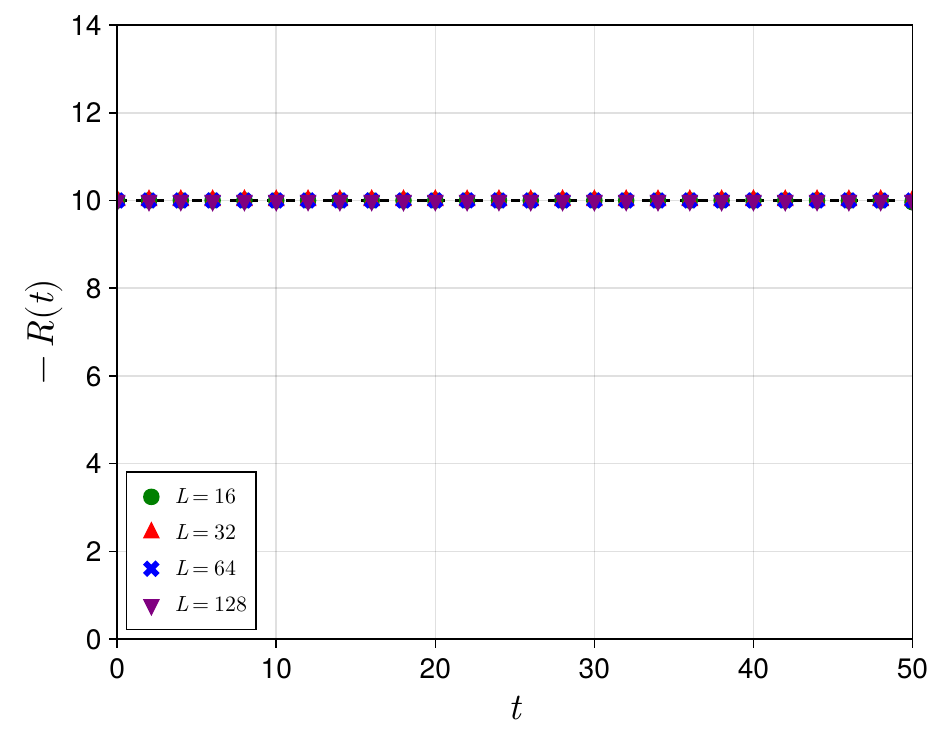}
    \caption{Results of $-R(t)$ from the mock data using our method with the quadratic function for $G(Q^2)$ with $g_1 = 0$ for $L = 16, 32, 64,$ and 128 with $A = 10$ as a function of $t$.}
    \label{fig:mock_our_method_L_dep_quad}
\end{figure}

In cases with other values of $A$ and $L$, which correspond to the analyses in the previous subsection presented in Figs.~\ref{fig:mock_imp_method_M_dep} and \ref{fig:mock_imp_method_L_dep}, our method using a quadratic function $G(Q^2)$ with $g_1 = 0$ works well as presented in Figs.~\ref{fig:mock_our_method_M_dep_quad} and \ref{fig:mock_our_method_L_dep_quad}.

\subsubsection{logarithmic function}

Another example of the choice of $G(Q^2)$ is a logarithmic function given by
\begin{equation}
    G(Q^2) = 1 + g_1^l \log ( 1 + g_2^l Q^2 ) .
\end{equation}
If $g_1^l$ and $g_2^l$ are adopted to satisfy $s_2 = 0$, the logarithmic function also can improve the convergence of $S(Q^2)$ compared to that of $F(Q^2)$.
This is expected from an estimate of the ratio of the expansion coefficients at the same order,
\begin{equation}
    \frac{s_n}{f_n} = \mathcal{O}\left(\left(\frac{g^l_2}{A}\right)^{2}
    \right) 
    \label{eq:our_method_log_sn}
\end{equation}
for $n \ge 3$ with $s_2 = 0$.
It can be smaller than unity when $g^l_2$ is smaller than $A$.
In this estimate we use $g_1^l = A^2/(g_2^l(A + g_2^l/2))$ determined from $s_2 = 0$ for a given $g_2^l$.

As in the previous example of the choice of $G(Q^2)$, the parameter set of $g_1^l$ and $g_2^l$ is determined from the value of $R_2(t)$.
In this analysis, we fix $g_2^l$ and investigate the $g_1^l$ dependence of $R_2(t)$.
In the following, we focus on the mock data with $A=10$ and $L=32$ as in the previous example.

\begin{figure}[ht!]
    \includegraphics*[scale=0.55]{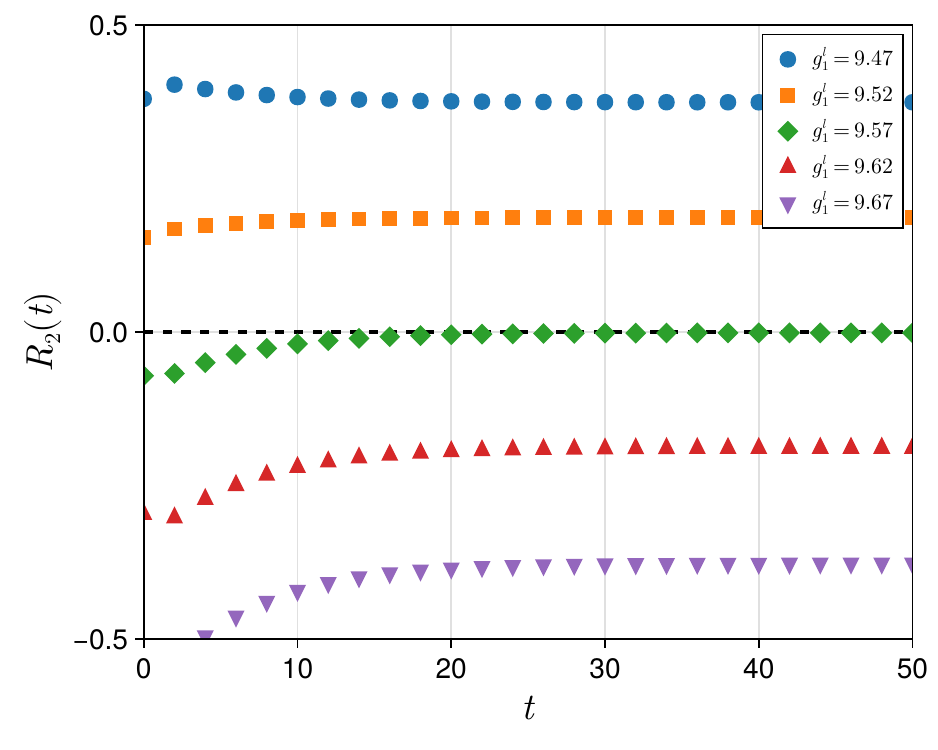}
    \caption{The same format as Fig.~\ref{fig:mock_our_method_1_quad}, but using the logarithmic function for several values of $g_1^l$ with fixed $g_2^l = 1$.}
    \label{fig:mock_our_method_1_log}
\end{figure}

Figure~\ref{fig:mock_our_method_1_log} presents the $t$ dependence of $R_2(t)$ with a logarithmic function with $g_2^l = 1$ and several values of $g_1^l$.
Although in a smaller $t$ region, $R_2(t)$ slightly depends on $t$, it is reasonably flat in a large-$t$ region.
Thus, we choose the same reference time, $t_r = 25$, as in the previous example to examine the $g_1^l$ dependence of $R_2(t)$.

\begin{figure}[ht!]
    \includegraphics*[scale=0.55]{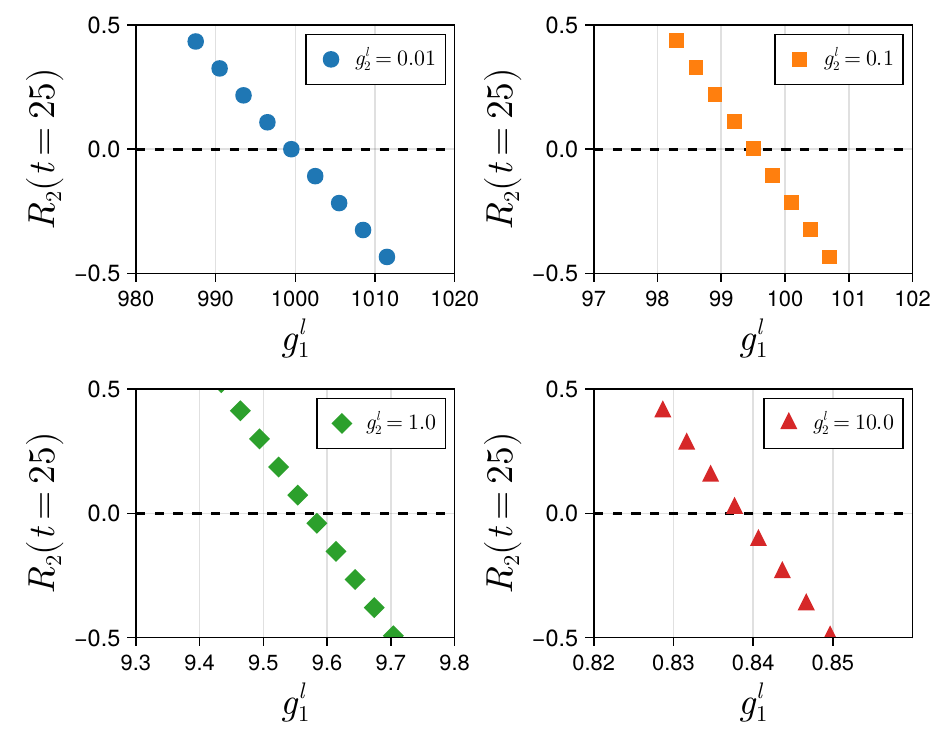}
    \caption{Dependence of $g_1^l$ near $R_2(t_r) = 0$ with several values of $g_2^l$ using the logarithmic function.}
    \label{fig:mock_our_method_2_log}
\end{figure}

The value of $R_2(t_r)$ is plotted in Fig.~\ref{fig:mock_our_method_2_log} as a function of $g_1^l$ in four $g_2^l$ cases.
As in the polynomial $G(Q^2)$ case, $R_2(t_r)$ reasonably behaves as a linear function of $g_1^l$ in each $g_2^l$ case, and the value of $g_2^l$ at $R_2(t_r)=0$ is easily obtained.
We note that $G(Q^2) \ne 0$ for all momenta in this case, because we consider only positive values for $g_1^l$ and $g_2^l$, so that $R_2(t)$ does not have a divergence in the logarithmic function.

\begin{figure}[ht!]
    \includegraphics*[scale=0.55]{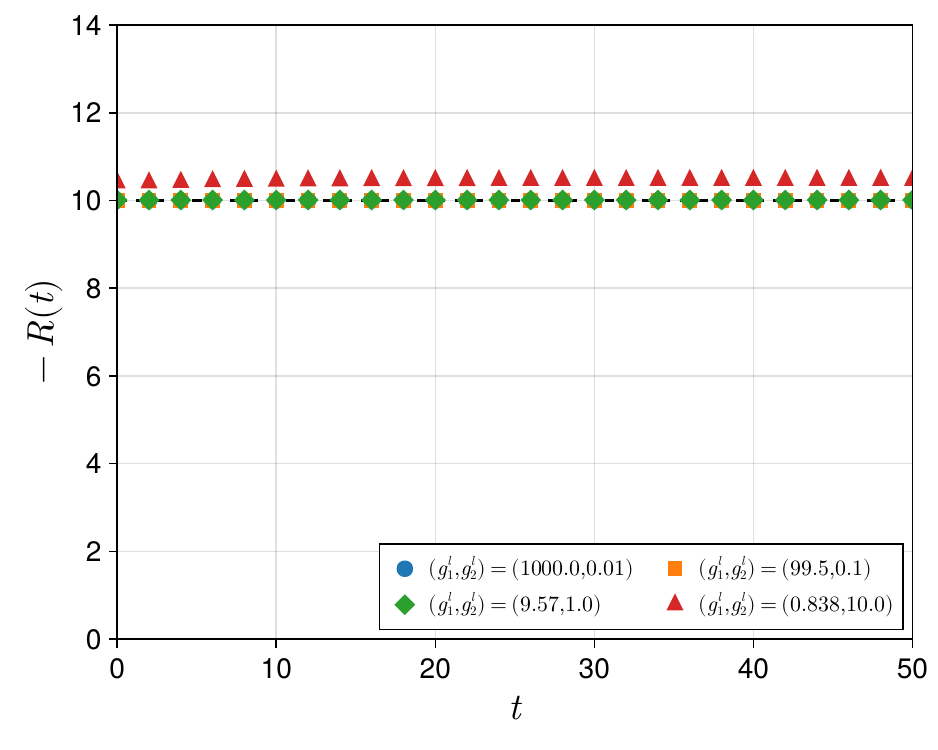}
    \caption{The same format as Fig.~\ref{fig:mock_our_method_4_quad}, but using the logarithmic function.}
    \label{fig:mock_our_method_3_log}
\end{figure}

\begin{figure}[ht!]
    \includegraphics*[scale=0.55]{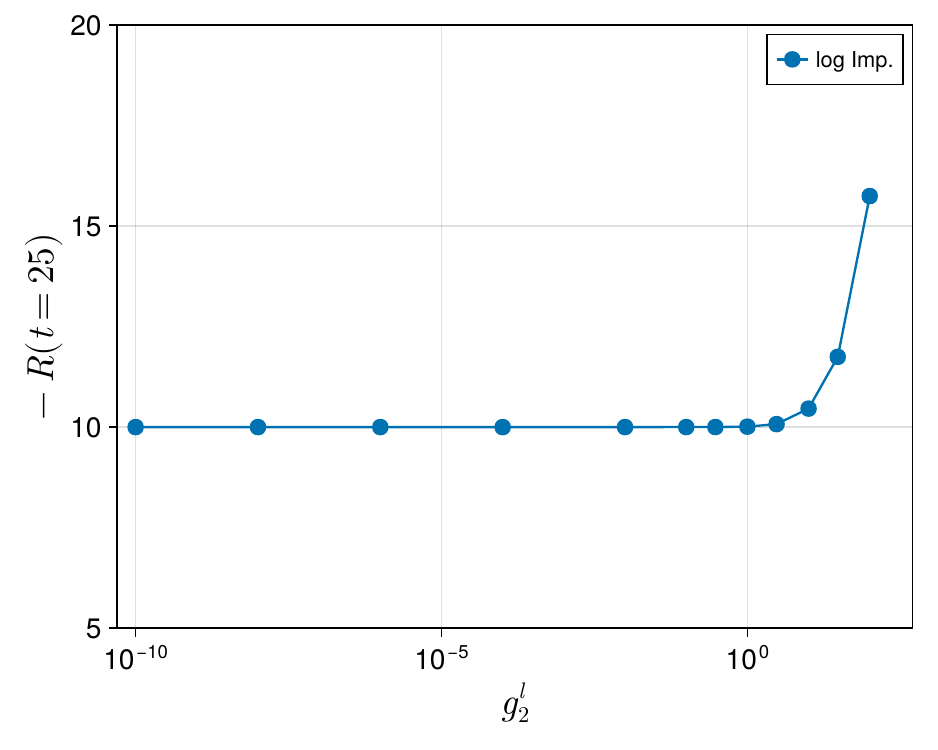}
    \caption{Result of $-R(t_r)$ as a function of $g_2^l$.}
    \label{fig:mock_our_method_4_log}
\end{figure}

The values of $-R(t)$ with the determined parameter sets behave as a constants as a function of $t$, as plotted in Fig.~\ref{fig:mock_our_method_3_log}.
While the result for a large $g_2^l$ differs from the input value of $A$ expressed by the dashed line, those for $g_2^l \le 1$ reasonably agree with the input.
The consistency for small $g_2^l$ is an expected tendency from Eq.~(\ref{eq:our_method_log_sn}).
The $g_2^l$ dependence of the result evaluated from $-R(t_r)$ is negligible for $g_2^l \le 1$ as shown in Fig.~\ref{fig:mock_our_method_4_log}.
From this trend, it is expected that the result with a logarithmic function is reasonably independent of $g_2^l$, when $g_2^l \le 1$, even in the large $A$ case.
This suggests that the analysis for the logarithmic function yields a stable result with respect to the choice of parameters in $G(Q^2)$.

\begin{figure}[ht!]
    \includegraphics*[scale=0.55]{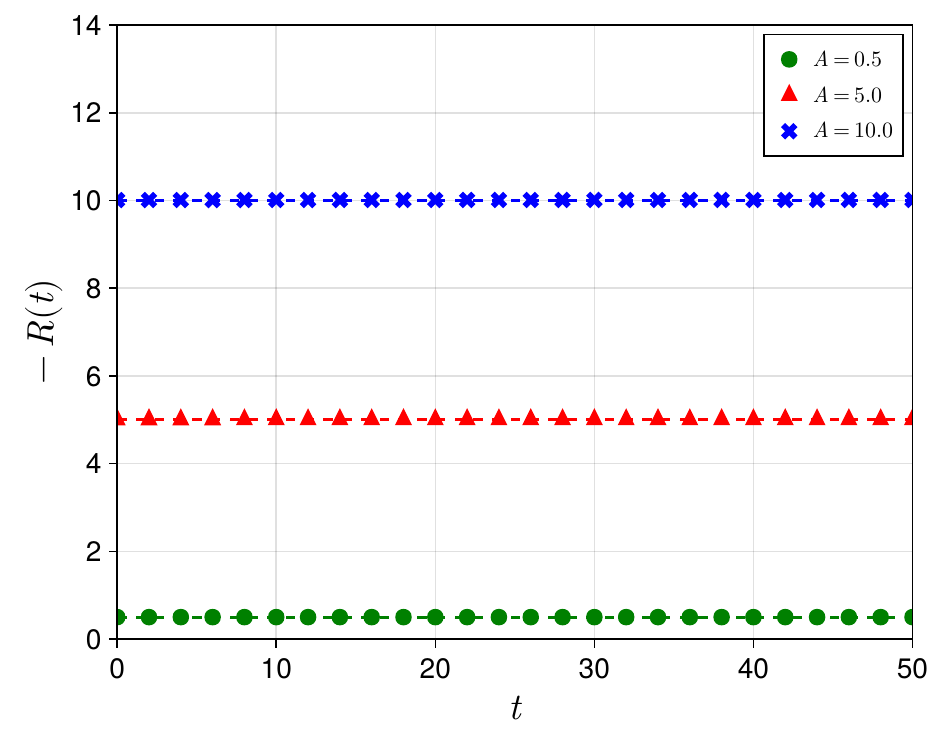}
    \caption{The same format as Fig.~\ref{fig:mock_our_method_M_dep_quad}, but using our method with the logarithmic function for $G(Q^2)$ with $g_2^l = 1$.}
    \label{fig:mock_our_method_M_dep_log}
\end{figure}

\begin{figure}[ht!]
    \includegraphics*[scale=0.55]{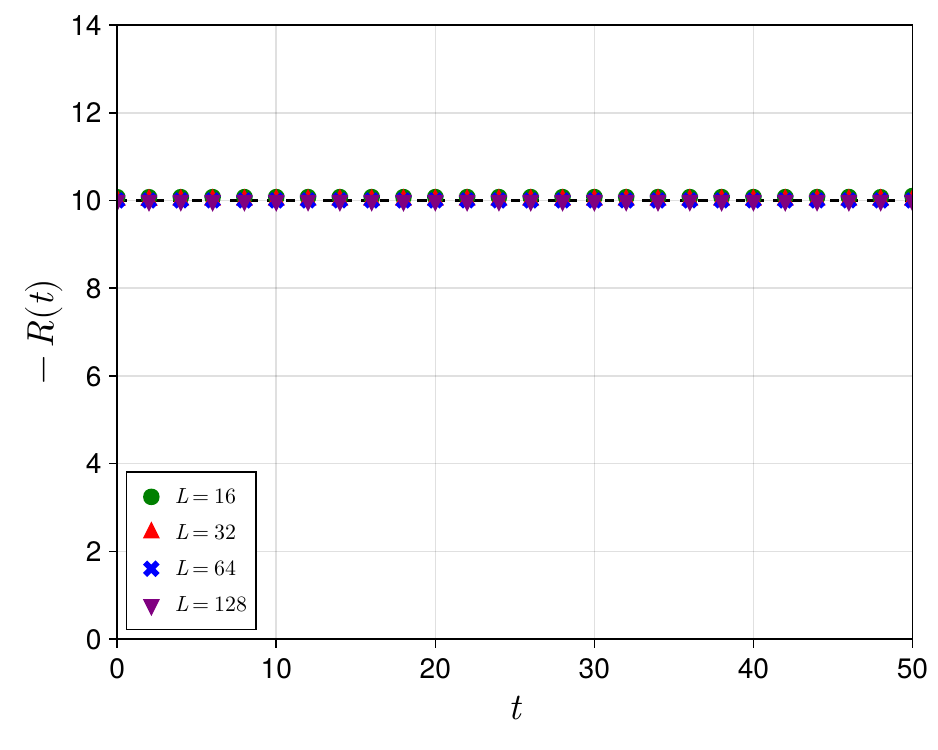}
    \caption{The same format as Fig.~\ref{fig:mock_our_method_L_dep_quad}, but using our method with the logarithmic function for $G(Q^2)$ with $g_2^l = 1$.}
    \label{fig:mock_our_method_L_dep_log}
\end{figure}

The mock data with several values of $A$ and $L$ as in Sec.~\ref{sec:mockup_improved_method} are analyzed with our method using the logarithmic function with $g_2^l = 1$.
Comparing with Figs.~\ref{fig:mock_imp_method_M_dep} and \ref{fig:mock_imp_method_L_dep}, the higher-order contributions are well suppressed in our method, and the results are consistent with the input $A$ expressed by the dashed lines in Figs.~\ref{fig:mock_our_method_M_dep_log} and \ref{fig:mock_our_method_L_dep_log} in all cases.

It is also noted that a $q$-logarithmic function defined by
\begin{equation}
    \ln_q(x) = \frac{x^{1-q}-1}{1-q} ,
\end{equation}
where $q \ne 1$, also could be a candidate for $G(Q^2)$ as
\begin{equation}
    G(Q^2) = 1 + g_1^q \ln_q(1+g_2^qQ^2),
\end{equation}
with parameters $g_1^q, g_2^q$, and $q$.
Similarly to the above examples of $G(Q^2)$, the higher-order contribution is expected to be suppressed when $s_2 = 0$ is satisfied in the small $g_2^q$ and $q$ case.
While we do not discuss the details, a similar result to that with the logarithmic function can be obtained with the $q$-logarithmic function.

\section{Results with lattice QCD data}
\label{sec:Numerical validation with actual lattice QCD configurations}

In this section, using actual lattice QCD data, we compare the results for the pion charge radius obtained from the linear combination and our methods as well as those obtained from fits to the form factor.

\subsection{Correlation functions}

The calculation of the moment of the 3-point function is basically the same as that for form factor calculations, for example, in Ref.~\cite{Ishikawa:2022ulx}.
To compute spatial moments of arbitrary order, we consider the 3-point function with one spatial coordinate defined as
\begin{equation}
    C_V(y_1-x_1,t) = 
    \sum_{\substack{x_2,x_3 \\ y_2,y_3}}
    \sum_{\vec{z}}
    \langle 0 | 
    O_\pi(\vec{z},t_f) V_4(\vec{y},t) O_\pi^\dagger(\vec{x},t_i) | 0 \rangle ,
    \label{eq:1d-3pt}
\end{equation}
where the pion field $O_\pi$ and the local electromagnetic current $V_4$ are defined by
\begin{eqnarray}
    O_\pi(x) &=& \overline{d}(x)\gamma_5 u(x)\\
    V_4(x) &=& \sum_{q=u,d,s} e_q \overline{q}(x)\gamma_4 q(x) 
\end{eqnarray}
with $e_u = 2/3$ and $e_d = e_s = -1/3$.
The source time $t_i$ is regarded as $0$ for simplicity in the following explanation, if $t_i$ appears.
The $2k$th-order moment of the 3-point function is calculated by
\begin{equation}
    C^{(k)}(t) = \sum_r r^{2k} C_V(r,t) ,
\end{equation}
with $-L/2+1 \le r \le L/2$.
In a middle-$t$ region far from both the source $t_i$ and sink $t_f$, where excited-state contributions are suppressed, the normalized moment $D^{(k)}(t) = C^{(k)}(t)/C^{(0)}(t)$ can be expressed by the pion electromagnetic form factor $F_\pi(Q^2)$ with the momentum transfer squared $Q^2 = 2m_\pi(E_\pi(p)-m_\pi)$ as
\begin{equation}
    D^{(k)}(t) = \sum_{p} R_Z(Q^2) F_\pi(Q^2) \Delta(p^2,t) T_k(p) ,
    \label{eq:1d-3pt_Dk}
\end{equation}
with $R_Z(Q^2) = Z_\pi(p)/Z_\pi(0)$, and the known functions $T_k(p)$ and $\Delta(p^2,t)$ defined in Eqs.~(\ref{eq:def_T_k}) and (\ref{eq:def_Delta_p_t}), respectively.
The amplitude $Z_\pi(p)$ is defined by $Z_\pi(p) = \langle 0 | O_\pi(0) | \pi(p) \rangle$ with $|\pi(p)\rangle$ being the pion state with momentum $p$.
This is because the 3-point function in the middle-$t$ region is represented by the form factor $F_\pi(Q^2)$ as
\begin{equation}
    C_V(r,t) = \frac{1}{Z_V}
    \sum_p \frac{Z_\pi(0) Z_\pi(p)}{2m_\pi} F_\pi(Q^2) \Delta(p^2,t)  
    e^{-m_\pi t_f}e^{-ipr} ,
\end{equation}
where $Z_V$ is the renormalization factor for the local vector current.

We evaluate the pion 2-point function with one spatial coordinate defined as
\begin{equation}
    C_\pi(y_1-x_1,t) = \sum_{\substack{x_2,x_3 \\ y_2,y_3}}
    \langle 0 | O_\pi(\vec{y},t) O_\pi^\dagger(\vec{x},t_i) | 0 \rangle .
    \label{eq:pion_2pt_x}
\end{equation}
Using $C_\pi(r,t)$, a momentum-projected 2-point function is calculated by
\begin{equation}
    \widetilde{C}_\pi(p^2,t) = \sum_r C_\pi(r,t) e^{ipr} .
    \label{eq:1d_pion_p_proj}
\end{equation}
From $\widetilde{C}_\pi(0,t)$ in a large-$t$ region, $m_\pi$ is determined from an exponential fit with 
\begin{equation}
    \widetilde{C}_\pi(0,t) = A \left(
    e^{-m_\pi t} + e^{-m_\pi (2T-t)}
    \right) ,
\end{equation}
where $A$ and $m_\pi$ are fit parameters, and $T$ is the temporal extent.
Note that the temporal extent is doubled in the fit function, because we employ an average of the correlators with periodic and anti-periodic boundary conditions in the temporal direction, which will be discussed below.

As a reference for the $\langle r_\pi^2 \rangle$ result, we also calculate the form factor $F_\pi(Q^2)$ itself at discrete $Q^2$, and obtain $\langle r_\pi^2 \rangle$ from a fit of $F_\pi(Q^2)$ with several fit-form assumptions.
Hereafter, we call this method with $F_\pi(Q^2)$ the traditional method.
In order to compute $F_\pi(Q^2)$, we measure the momentum-projected 3-point function with $\vec{p}$ defined by
\begin{equation}
    \widetilde{C}_V^{\rm 3d}(p^2,t) = 
    \sum_{\vec{x},\vec{y},\vec{z}}
    \langle 0 | 
    O_\pi(\vec{z},t_f) V_4(\vec{y},t) O_\pi^\dagger(\vec{x},t_i) | 0 \rangle 
    e^{i\vec{p}(\vec{y}-\vec{x})} .
    \label{eq:3d-3pt}
\end{equation}
The renormalized $F_\pi(Q^2)$ is obtained from an appropriate ratio of $\widetilde{C}_V^{\rm 3d}(p^2,t)$ given by
\begin{equation}
    F_\pi(Q^2) = \frac{1}{R_Z^{\rm 3d}(Q^2)}\frac{2E_\pi(\vec{p})}{E_\pi(\vec{p})+m_\pi}
    \frac{\widetilde{C}_V^{\rm 3d}(p^2,t)}{\widetilde{C}_V^{\rm 3d}(0,t)} e^{(E_\pi(\vec{p})-m_\pi)t}
    \label{eq:3d-FormFactor}
\end{equation}
in the middle-$t$ region between the source $t_i$ and sink $t_f$.

We determine the ratio of amplitudes $R_Z^{\rm 3d}(Q^2)$ for each momentum transfer squared $Q^2$ from the momentum-projected 2-point function $\widetilde{C}_\pi^{\rm 3d}(p^2,t)$, which is evaluated as
\begin{equation}
    \widetilde{C}_\pi^{\rm 3d}(p^2, t) =
    \sum_{\vec{x},\vec{y}}
    \langle 0 |
    O_\pi(\vec{y},t)O_\pi^\dagger(\vec{x},t_i) | 0 \rangle
    e^{i\vec{p}(\vec{y}-\vec{x})}\, .
\end{equation}
$R_Z^{\rm 3d}(Q^2)$ is constructed as
\begin{equation}
    R_Z^{\rm 3d}(Q^2) = \sqrt{
    \frac{E_\pi(\vec{p})}{m_\pi}
    \frac{e^{-m_\pi t} + e^{-m_\pi(2T-t)}}{e^{-E_\pi(\vec{p})t} + e^{-E_\pi(\vec{p})(2T-t)}}
    \frac{\widetilde{C}_\pi^{\rm 3d}(p^2,t)}{\widetilde{C}_\pi^{\rm 3d}(0,t)}
    }\, ,
    \label{eq:meas_R_Z_q2}
\end{equation}
and is extracted from a plateau in a sufficiently large-$t$ region, where the excited-state contaminations are suppressed.

The mean square pion charge radius is defined by the slope of $F_\pi(Q^2)$ as
\begin{equation}
    \langle r_\pi^2 \rangle = - 6 \left.\frac{d F_\pi(Q^2)}{dQ^2}\right|_{Q^2=0} .
\end{equation}
We will compare the results for $\langle r_\pi^2 \rangle$ obtained from the linear combination and our methods as well as the traditional method below.

\subsection{Simulation parameters}

For this investigation, $N_f = 2+1$ gauge ensembles are adopted, which were generated by the PACS-CS Collaboration using the Iwasaki gauge action and the nonperturbative $O(a)$-improved Wilson quark action at $m_\pi = 0.51$ GeV.
The details of the gauge ensembles are described in Ref.~\cite{Yamazaki:2012hi}.
The lattice sizes are $L^3\times T = 32^3 \times 48$ and $64^3\times 64$.
Using the gauge ensembles, we measure the 2- and 3-point functions explained above with the same quark action as that used for the gauge ensemble generation.
Parameters for the ensemble are tabulated in Table~\ref{tab:sim_para}.
We also calculate the charge radius using the configurations at $m_\pi=0.3$ GeV, whose results are presented in a later subsection and in Appendix~\ref{app:result_mpi0.3}.

For the calculation of the 3-point functions, the Z(2) $\otimes$ Z(2) random source~\cite{Boyle:2008yd} is utilized on the source time slice $t_i$ to reduce the calculation cost.
For the 3-point function in Eq.~(\ref{eq:1d-3pt}), the random number is spread in the $x_2$-$x_3$ plane at $t_i$ as well as over the color and spin spaces.
The same random source is employed in the calculation of the pion 2-point function in Eq.~(\ref{eq:pion_2pt_x}).
On the other hand, for the momentum-projected 3-point function in Eq.~(\ref{eq:3d-3pt}), we utilize the random source spread over the spatial volume and the color and spin spaces, which is the same as in the $K_{\ell 3}$ form factor calculations~\cite{PACS:2019hxd,Ishikawa:2022ulx}.
We also employ the sequential source method~\cite{Kilcup:1985fq,Martinelli:1988rr} at the sink time slice $t_f$ for cost reduction.
In both calculations for the 3-point functions, we fix $|t_f-t_i| = 22$.

We compute the quark propagators with both periodic and anti-periodic boundary conditions in the temporal direction and take the average of the resulting correlators.
This average suppresses contributions in which intermediate states wrap around the temporal boundary in the 3-point functions, called the wrapping-around effect~\cite{Aoki:2008ss,Kakazu:2017fhv,PACS:2019hxd,Ishikawa:2022ulx}, and it is equivalent to working with an effective temporal extent of $2T$ for 2-point functions. 
Accordingly, unless explicitly stated otherwise, all correlators in this study are understood to be averages over the periodic and anti-periodic boundary conditions.
For the 3-point functions we further exploit the fact that, due to the temporal periodicity, the pion can propagate not only forward from the source to the sink but also in the opposite time direction, {\it i.e.}, from the source to the sink around the temporal boundary~\cite{QCDSFUKQCD:2006gmg}.
For example, with $t_i=0$, we average the data between the forward-propagating 3-point function with the sink fixed at $t_f$ and the backward-propagating 3-point function with the sink fixed at $T-t_f$.
This provides an additional factor-of-two statistical average.

In order to increase the number of measurements of $C_V(r,t)$, we repeat the calculations with different source time slices $t_i$ and spatial points $x_1$ in Eq.~(\ref{eq:1d-3pt}) on each gauge configuration.
Furthermore, we measure the 3-point function $C_V(r,t)$ in each spatial direction to improve its signal.
The numbers of the measurements are summarized in Table~\ref{tab:meas_para}.
For the momentum-projected 3-point function, we employ $0 \le p^2 \le (2\pi/L)^2 \cdot 6$, and the number of its measurements on each configuration is also shown in Table~\ref{tab:meas_para}.

\begin{table}[t!]
    \caption{Simulation parameters for the 2 + 1 flavor gauge ensembles.
    $L, T, a$ and $N_{\rm conf}$ represent the spatial and temporal extents, the lattice spacing, and the number of configurations, respectively.
    $m_\pi$ is the measured pion mass.}
    \label{tab:sim_para}
    \begin{tabular}{cccccccc}
        \hline\hline
        $\beta$ & $L^3\times T$ & $a^{-1}$[GeV] & $a$[fm] & $L$[fm] & $m_\pi$[GeV] & $N_{\rm conf}$ \\\hline
        1.90 & 32$^3\times$48 & 2.194(10) & 0.08995(40) & 2.9 & 0.511(3)  & 80 \\
        1.90 & 64$^3\times$64 & 2.194(10) & 0.08995(40) & 5.8 & 0.511(2)  & 54 \\
        \hline\hline
    \end{tabular}
\end{table}

\begin{table}[t!]
    \caption{Details of the measurements for the one- and three-dimensional correlation functions: the number of source points and random sources ($N_{\rm src}$ and $N_{\rm r}$), and the number of measurements per configuration ($N_{\rm meas}$) are summarized.
    For $N_{\rm meas}$, in the one-dimensional case it is calculated as $N_{\rm meas} = 12 \times N_{\rm src} \times N_{\rm r}$, while in the three-dimensional case it is $N_{\rm meas} = 4 \times N_{\rm src} \times N_{\rm r} \times N_{d}$, where $N_{d}$ is the degeneracy of momentum vectors with the same magnitude in three-dimensional space.
    In this study, $N_{d}$ takes the values $1, 6, 12, 8, 6, 9, 9$ as the momentum magnitude increases from zero. The two largest values would normally be 24, but not all cases were computed to reduce computational cost.}
    \label{tab:meas_para}
    \begin{tabular}{cccccccc}
        \hline\hline
        Type of data & $L^3\times T$ & $N_{\rm src}$ & $N_{\rm r}$ & $N_{\rm meas}$ \\\hline
        1d & 32$^3\times$48 & 16 & 4 & 768 \\
           & 64$^3\times$64 & 32 & 1 & 384 \\\hline
        3d & 32$^3\times$48 & 4 & 32 & $512 \times N_{d}$ \\
           & 64$^3\times$64 & 4 & 6  &  $96 \times N_{d}$ \\
        \hline\hline
    \end{tabular}
\end{table}

\subsection{Traditional method}

\begin{figure}[ht!]
    \includegraphics[keepaspectratio, scale=1.0]{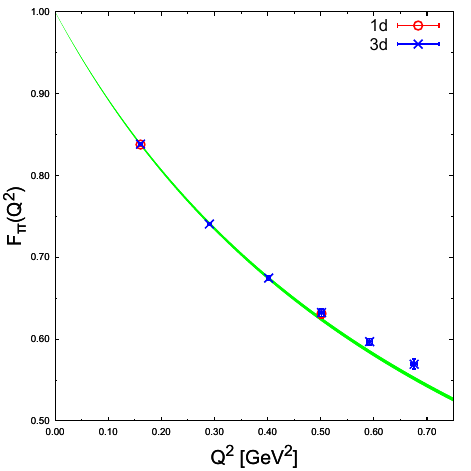}
    \caption{Result of $F_\pi(Q^2)$ at $m_\pi=0.5$ GeV on $L = 32$ as a function of $Q^2$.
    The green curve denotes the result of a monopole fit.
    The blue symbols represent data points obtained from Eq.~(\ref{eq:3d-FormFactor}), while the red symbols are calculated in the same manner using the 3-point function in one spatial coordinate.}
    \label{fig:ff_pi_mpi0.5_L32}
\end{figure}

The measured $F_\pi(Q^2)$ on $L=32$ is presented in Fig.~\ref{fig:ff_pi_mpi0.5_L32} as a function of $Q^2$.
The values of $F_\pi(Q^2)$ are presented in Table~\ref{tab:ff_mpi0.5_L32}.
The data for $F_\pi(Q^2)$ are fitted by three fit forms: monopole form, cubic forms of $Q^2$, and $z$-parameter expansion~\cite{Lee:2015jqa} given by
\begin{equation}
    F_\pi(Q^2) = \left\{
    \begin{array}{l}
        \left(1+AQ^2\right)^{-1}\\
        1 - A Q^2 + B Q^4 + C Q^6\\
        1 - a_1 z + a_2 z^2 + a_3 z^3
    \end{array}
    \right.\ ,
\end{equation}
where $A,B,C$ and $a_1,a_2,a_3$ are fit parameters.
In the $z$-parameter expansion, we follow the definition of the $z$-parameter in Ref.~\cite{Gao:2021xsm}, with $t_0 = 0$ for simplicity as
\begin{equation}
    z = \frac{\sqrt{Q^2 + t_{\rm cut}}-\sqrt{t_{\rm cut}}}
    {\sqrt{Q^2 + t_{\rm cut}}+\sqrt{t_{\rm cut}}}
    \label{eq:z-para}
\end{equation}
with $t_{\rm cut} = 4m_\pi^2$.
From the fit results, $\langle r^2_\pi \rangle$ is determined through $\langle r^2_\pi \rangle = 6A$ or $3 a_1 / (2 t_{\rm cut})$.
The fit result with the monopole form is plotted in Fig.~\ref{fig:ff_pi_mpi0.5_L32} as a typical result.
The fit results for $\langle r^2_\pi \rangle$ are tabulated in Table~\ref{tab:trad_r2_fit_results_L32} together with the uncorrelated $\chi^2$/d.o.f. for each fit.
For the fit ranges, we use all six $Q^2$ data points for the cubic-polynomial fit and the $z$-expansion fit, while the monopole fit is performed using only the three lowest $Q^2$ points.
This is because it is difficult to obtain a reasonable value of $\chi^2/$d.o.f. over a wider range of $Q^2$ with the monopole fit.

\begin{table}[t!]
    \caption{Result of $F_\pi(Q^2)$ for each $Q^2$ at $m_\pi=0.5$ GeV on $L = 32$.}
    \label{tab:ff_mpi0.5_L32}
    \begin{tabular}{cc}
        \hline
        \hline
        $Q^2$[GeV$^2$] & $F_\pi(Q^2)$ \\
        \hline
        0.161(1) & 0.838(1) \\
        0.290(3) & 0.741(2) \\
        0.402(4) & 0.675(3) \\
        0.501(5) & 0.632(5) \\
        0.592(6) & 0.597(4) \\
        0.676(6) & 0.569(6) \\
        \hline
        \hline
    \end{tabular}
\end{table}

\begin{table}[t!]
    \caption{Results of $\langle r^2_\pi \rangle$ obtained from a fit to $F_\pi(Q^2)$ with each fit form and their weighted average at $m_\pi=0.5$ GeV on $L = 32$.}
    \label{tab:trad_r2_fit_results_L32}
    \begin{tabular}{ccc}
        \hline
        \hline
        fit form  & $\langle r^2_\pi \rangle$[fm$^2$] & $\chi^2$/d.o.f. \\
        \hline
        monopole  & 0.281(4) & 0.1 \\
        cubic     & 0.272(3) & 0.5 \\
        $z$-exp   & 0.274(4) & 0.7 \\
        average   & $0.276(2)_{\rm stat.}(5)_{\rm sys.}$ & - \\
        \hline
        \hline
    \end{tabular}
\end{table}

\begin{figure}[ht!]
    \includegraphics*[scale=1.0]{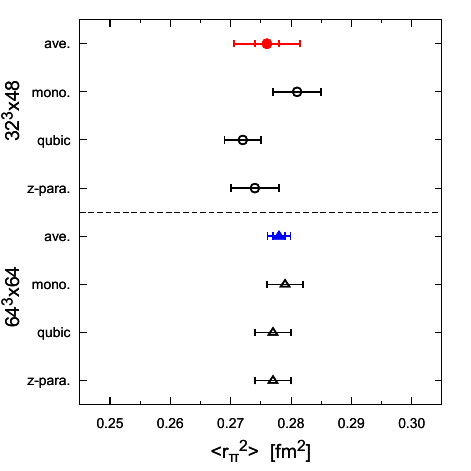}
    \caption{Results of $\langle r^2_\pi \rangle$ obtained from a fit to $F_\pi(Q^2)$ with each fit form and their weighted average at $m_\pi=0.5$ GeV.
    The result of a specific fit form is represented by the open symbol.
    The filled symbols are the weighted average of the three results.
    The inner (outer) error of the filled symbols represents the statistical error (combined error with the statistical and systematic errors added in quadrature).}
    \label{fig:trad_r2_mpi0.5}
\end{figure}

The results from each fit are summarized in Fig.~\ref{fig:trad_r2_mpi0.5}.
Since a fit-form dependence is observed, we estimate a systematic error associated with this dependence.
To do this, we first determine the central value by a weighted average of the three results.
Its statistical error is evaluated by the jackknife method.
A systematic error coming from the fit-form dependence is estimated from the maximum difference between the individual fit results and the central value.
We regard the central value and the statistical and systematic errors as the result from the traditional method on this volume, which is presented by the filled circle in Fig.~\ref{fig:trad_r2_mpi0.5}.

\begin{figure}[ht!]
    \includegraphics[keepaspectratio, scale=1.0]{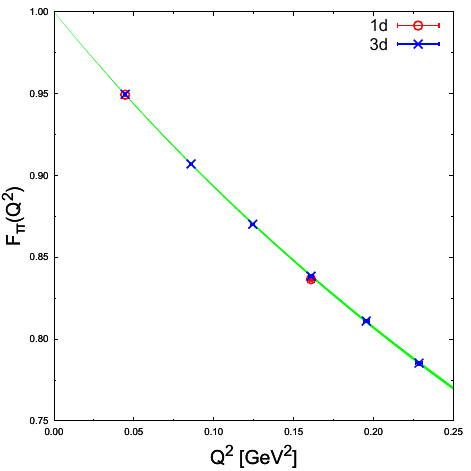}
    \caption{The same format as Fig.~\ref{fig:ff_pi_mpi0.5_L32}, but for $L = 64$.}
    \label{fig:ff_pi_mpi0.5_L64}
\end{figure}

The same calculation is performed on the $L=64$ lattice.
The data of $F_\pi(Q^2)$, whose values are tabulated in Table~\ref{tab:ff_mpi0.5_L64}, and the fit result with the monopole form are presented in Fig.~\ref{fig:ff_pi_mpi0.5_L64}.
The results for $\langle r^2_\pi \rangle$ obtained from the three fits are summarized in Table~\ref{tab:trad_r2_fit_results_L64}.
For the fit ranges on $L=64$ lattice, we use all six $Q^2$ data points for all fits.
The results of $\langle r^2_\pi \rangle$ are summarized in Fig.~\ref{fig:trad_r2_mpi0.5}, which shows that the fit-form dependence becomes milder than that on $L=32$.
This can be interpreted as follows: the values of $F_\pi(Q^2)$ at small $Q^2$ are obtained in the small unit of the lattice momentum, $2\pi/L$, and thus the fit of $F_\pi(Q^2)$ becomes stable in determining the slope at the origin.
The central value with the systematic error on $L=64$ obtained from
the three fit results is plotted in the same figure.

\begin{table}[t!]
    \caption{The same format as Table~\ref{tab:ff_mpi0.5_L32}, but for $L = 64$.}
    \label{tab:ff_mpi0.5_L64}
    \begin{tabular}{cc}
        \hline
        \hline
        $Q^2$[GeV$^2$] & $F_\pi(Q^2)$ \\
        \hline
        0.0445(04) & 0.9497(02) \\
        0.0857(08) & 0.9071(04) \\
        0.1244(11) & 0.8702(05) \\
        0.1608(15) & 0.8385(08) \\
        0.1954(18) & 0.8111(09) \\
        0.2284(21) & 0.7853(11) \\
        \hline
        \hline
    \end{tabular}
\end{table}

\begin{table}[t!]
    \caption{The same format as Table~\ref{tab:trad_r2_fit_results_L32}, but for $L = 64$.}
    \label{tab:trad_r2_fit_results_L64}
    \begin{tabular}{ccc}
        \hline
        \hline
        fit form  & $\langle r^2_\pi \rangle$[fm$^2$] & $\chi^2$/d.o.f. \\
        \hline
        monopole  & 0.279(3) & 0.3 \\
        cubic     & 0.277(3) & 0.6 \\
        $z$-exp   & 0.277(3) & 0.6 \\
        average   & $0.278(1)_{\text{stat.}}(1)_{\text{sys.}}$ & - \\
        \hline
        \hline
    \end{tabular}
\end{table}

\subsection{Model-independent methods}

In this subsection, we compare the results obtained from the linear combination and our methods.

\subsubsection{Analysis of Amplitude Ratio}

\begin{figure}[ht!]
    \includegraphics*[scale=0.8]{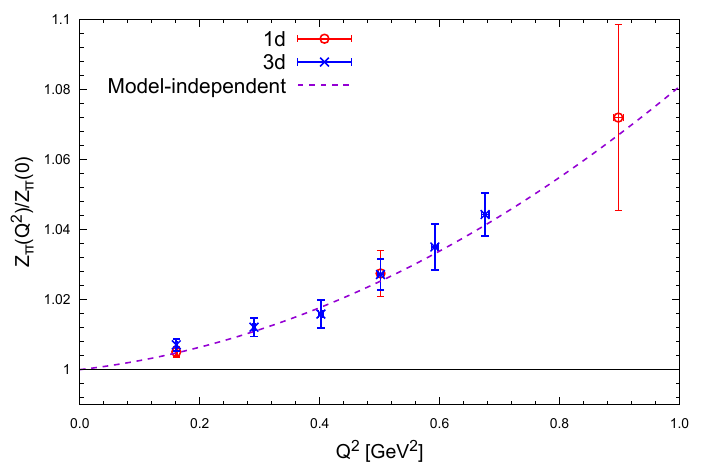}
    \caption{Data of $R_Z(Q^2)$ calculated from the momentum-projected 2-point function and the function of $R_Z(Q^2)$ determined from a model-independent method.
    The purple dashed line is drawn using parameters obtained by extracting $Z_1$ and $Z_2$ in the linear combination method.}
    \label{fig:model-indep_Zpi_ov_Zp_mpi0.5_L32}
\end{figure}

Although the ratio $R_Z(Q^2)=Z_\pi(p)/Z_\pi(0)$ should be unity in the continuum theory, it is not exactly unity at finite lattice spacing.
In fact, our data show a slight deviation from unity, as seen in Fig.~\ref{fig:model-indep_Zpi_ov_Zp_mpi0.5_L32}.
Since correctly evaluating the effects originating from $R_Z(Q^2)$ is important for a reliable determination of the charge radius, one would ideally evaluate $R_Z(Q^2)$ for all possible momenta $p$ appearing in Eq.~(\ref{eq:1d-3pt_Dk}). 
In practice, however, this is difficult.
Instead, we apply the model-independent method to quantities in which $R_Z(Q^2)$ is included multiplicatively.
Specifically, in the linear combination method we analyze $F_\pi(Q^2)R_Z(Q^2)$, while in our method we analyze $F_\pi(Q^2)G(Q^2)R_Z(Q^2)$.
In this way, the first-order coefficients in the Taylor expansion of these products at $Q^2=0$ can be extracted directly from the spatial moments, without requiring an explicit determination of $R_Z(Q^2)$ at each momentum.

To obtain the slope of the form factor $F_\pi(Q^2)$, we subtract the contribution of $R_Z(Q^2)$ afterward.
We expand $R_Z(Q^2)$ as
\begin{equation}
    R_Z(Q^2)=1+Z_1 Q^2+Z_2 Q^4+\cdots ,
    \label{eq:def_Z_i}
\end{equation}
and determine the coefficient $Z_1$ using a model-independent analysis, in close analogy to the determination of $\langle r_\pi^2\rangle$.
In particular, we employ the linear combination method applied to the moments of the 2-point function $C_\pi(r,t)$ defined in Eq.~(\ref{eq:pion_2pt_x}).
That is, using the normalized moments of the 2-point function,
\begin{equation}
    D^{(k)}_\pi(t) = \frac{\sum_r r^{2k} C_{\pi}(r,t)}{\sum_r C_{\pi}(r,t)} ,
\end{equation}
we construct the quantity defined in Eq.~(\ref{eq:imp_method_def_R}).
The parameters $\overline{\alpha}_0$, $\overline{\alpha}_1$, and $\overline{\alpha}_2$ are determined so as to satisfy Eqs.~(\ref{eq:condition_1})--(\ref{eq:condition_3}).
Here, $\overline{\beta}_{nk}(t)$ appearing in Eq.~(\ref{eq:def_beta-bar}) is replaced by
\begin{equation}
    \overline{\beta}_{nk}^\pi(t) = \sum_p Q^{2n} \Delta_\pi(p^2,t) T_k(p) ,
\end{equation}
where $\Delta_\pi(p^2,t)$ is defined as
\begin{equation}
    \Delta_\pi(p^2,t) =
    \dfrac{m_{\pi}}{E_{\pi}(p)}
    \dfrac{e^{-E_{\pi}(p)t} + e^{-E_{\pi}(p)(2T-t)}}
    {e^{-m_{\pi}t} + e^{-m_{\pi}(2T-t)}} .
\end{equation}
With this set of parameters, the counterpart of $\overline{R}(t)$ in Eq.~(\ref{eq:imp_method_R}) behaves as
\begin{equation}
    \overline{R}_\pi(t) = Z_1
    + \sum_{n\ge3} Z_n \left(
    \overline{\alpha}_1\overline{\beta}_{n1}^\pi(t)
    + \overline{\alpha}_2\overline{\beta}_{n2}^\pi(t)
    \right) .
\end{equation}

Figure~\ref{fig:model-indep_Zpi_ov_Zp_mpi0.5_L32} shows the consistency of the model-independent analysis with the measured data determined from the ratio in Eq.~(\ref{eq:meas_R_Z_q2}).
Since $R_Z(0)=1$ by definition, the desired first Taylor coefficient of
$F_\pi(Q^2)$ is then obtained by subtracting $Z_1$ from the slope extracted for $F_\pi(Q^2)R_Z(Q^2)$, {\it i.e.,} $\overline{R}(t)-Z_1$ (or, equivalently, from that extracted for $F_\pi(Q^2)G(Q^2)R_Z(Q^2)$).

\subsubsection{Analysis of charge radius}
\label{sec:result:ana_L32_m0.5:r2}

\begin{figure}[ht!]
    \includegraphics*[scale=0.8]{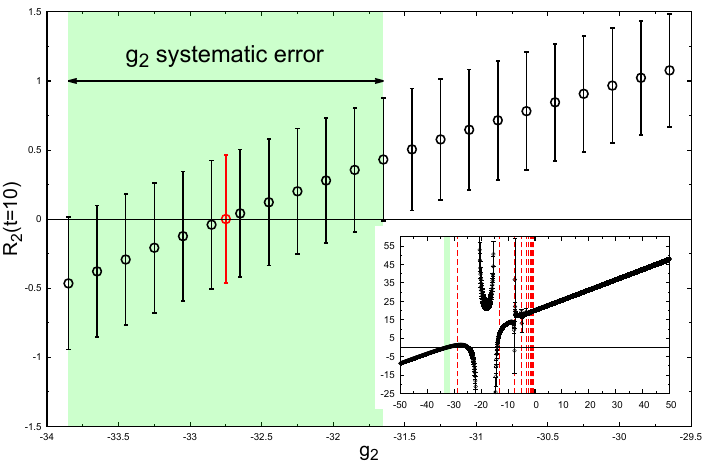}
    \caption{Data of $R_2(t)$ for the quadratic function with $g_1=0$ as a function of $g_2$ at $m_\pi=0.5$ GeV on $L=32$.
    The dashed lines express the values of $g_2$ for $G(Q^2) = 0$.
    }
    \label{fig:our_method_quad_R2t_mpi0.5_L32}
\end{figure}

In our method, we employ a quadratic function $G(Q^2) = 1 + g_1 Q^2 + g_2 Q^4$ and a logarithmic function $G(Q^2) = 1 + g_1^l \log( 1 + g_2^l Q^2 )$, as discussed in the previous section.
In both cases, the parameter sets are determined from $R_2(t) = 0$ defined in Eq.~(\ref{eq:def_R2t}).
An example of the data for $R_2(t) = 0$ with the quadratic function is plotted in Fig.~\ref{fig:our_method_quad_R2t_mpi0.5_L32}.
From the $g_2$ dependence of $R_2(t)$, we determine the value of $g_2$ satisfying $R_2(t) = 0$.
In this determination we neglect $R_2(t)=0$ near a region with $G(Q^2) = 0$, which may cause a singular behavior of $R_2(t)$, as discussed in Sec.~\ref{sec:method:our_method}.

\begin{figure}[ht!]
    \includegraphics*[scale=0.8]{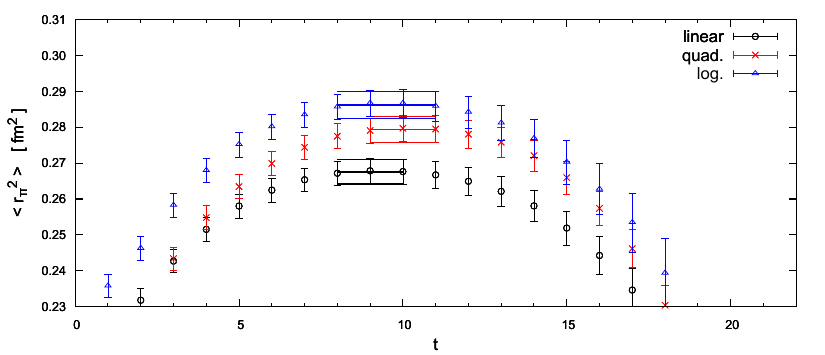}
    \caption{The time dependence of $-6(R(t)-Z_1)$ with Eq.~(\ref{eq:our_method_def_Rt}) or $-6(\overline{R}(t)-Z_1)$ with Eq.~(\ref{eq:imp_method_def_R}) at $m_\pi=0.5$ GeV on $L=32$.  The labels ``linear'', ``quad.'', and ``log.'' represent the linear combination method and our improved methods with the quadratic and logarithmic functions, respectively.
    For each solid line, the range and the width represent, respectively, the fit range of the constant fit and the central value of the extracted result with its statistical error.}
    \label{fig:model-indep_Rt_mpi0.5_L32}
\end{figure}

Figure~\ref{fig:model-indep_Rt_mpi0.5_L32} plots the data for $-6(\overline{R}(t)-Z_1)$ in Eq.~(\ref{eq:imp_method_def_R}) and $-6(R(t)-Z_1)$ in Eq.~(\ref{eq:our_method_def_Rt}) as a function of $t$: the former is evaluated by the linear combination method, while the latter is done by using the quadratic function with $g_1 = 0$ and the logarithmic function with $g_2^l = 1$.
The results are consistent with each other near the source and sink time slices, $t=0$ and $t=22$, respectively.
In contrast to these regions, the data from the linear combination method is about 5\% smaller than those of our method in the middle-$t$ region.
We determine the values of $\langle r_\pi^2 \rangle$ from a constant fit in the middle-$t$ region.

\begin{figure}[ht!]
    \includegraphics*[scale=0.8]{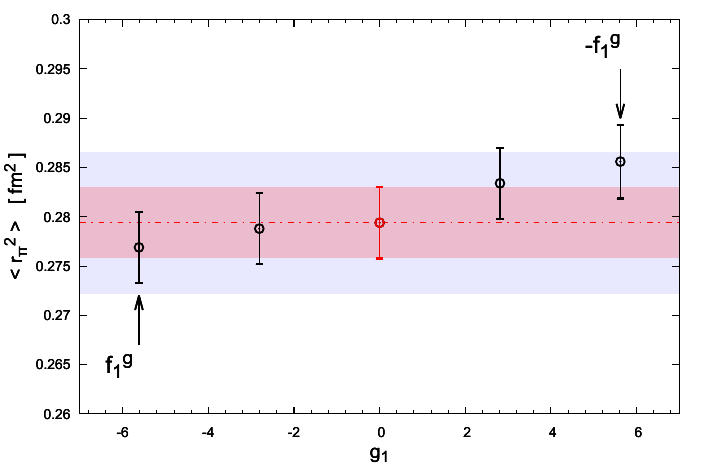}
    \caption{Dependence of the charge radius on the parameter $g_1$ for the quadratic function.
    The red symbol is adopted as the central value with statistical errors, and the maximum deviation from it is evaluated as the systematic error.
    The red band indicates only the statistical error, while the blue band shows the uncertainty range including the systematic error of this $g_1$.
    }
    \label{fig:our_method_quad_sys_mpi0.5_L32}
\end{figure}

In our method, a systematic error arising from the parameter choices of each $G(Q^2)$ needs to be evaluated.
For the quadratic function, we simply employ the analysis with $g_1 = 0$ as our reference.
We note that $g_1 = -\langle r^2_\pi \rangle/6$ is an alternative choice since it is expected that $F_\pi(Q^2)$ at lower $Q^2$ is expressed by a monopole form.
The systematic error is estimated from the maximum difference between the central values with $g_1 = 0$ and that with $g_1=\pm f_1^g$, where $f_1^g$ is determined from the central value of the weighted average of the traditional method.
The results with the three choices of $g_1$ are plotted in Fig.~\ref{fig:our_method_quad_sys_mpi0.5_L32} together with the ones with $g_1 = \pm 0.5f_1^g$.
While the results have a mild linear dependence on $g_1$, this dependence gives the largest systematic error in the analysis with the quadratic function.

We estimate the systematic error associated with $g_2$ in the quadratic choice for $G(Q^2)$ as follows.
We fix $g_1=0$ and determine $g_2$ by scanning $g_2$ and searching for the solution of $R_2(t)=0$ as shown in Fig.~\ref{fig:our_method_quad_R2t_mpi0.5_L32}: the red symbol indicates the central $g_2$ value, while the green band represents the range of $g_2$ for which $R_2(t)=0$ is satisfied within the $1\sigma$ statistical uncertainty.
Due to the statistical uncertainty of $R_2(t)$, the condition $R_2(t)=0$ does not determine a unique value of $g_2$.
Instead, there exists a finite range of $g_2$ values for which $R_2(t)$ is consistent with zero within its $1\sigma$ statistical error.
We define the central value of $g_2$ as the point where the central value of $R_2(t)$ satisfies $R_2(t)=0$.
The systematic error from $g_2$ is then estimated in close analogy to the procedure for the $g_1$ dependence.
By varying $g_2$ within the allowed range, we take the maximum deviation of the final result (e.g. $\langle r_\pi^2\rangle$) from the central value as the systematic error induced by $g_2$.

\begin{figure}[ht!]
    \includegraphics*[scale=0.8]{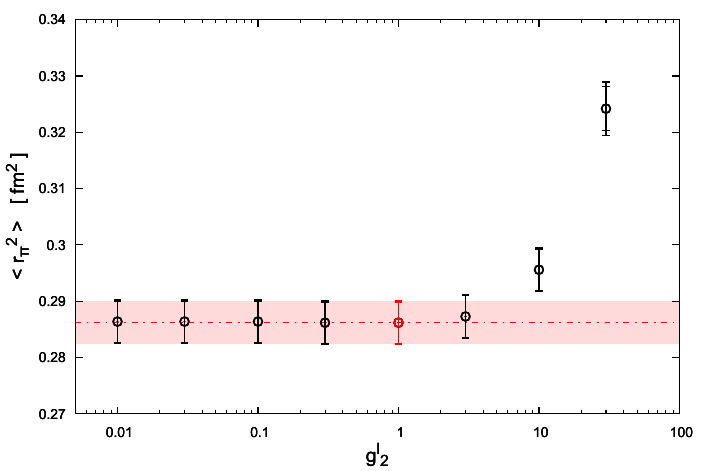}
    \caption{Dependence of the charge radius on the parameter $g_2^l$ when a logarithmic function is chosen for $G(Q^2)$.
    The data shown by the red symbol is adopted as the central value with statistical uncertainties, and the maximum deviation of the results obtained in the region $g_2^l < 1$ from it is taken as the systematic uncertainty.
    The red band indicates the range of uncertainty due to statistical errors only.}
    \label{fig:our_method_log_sys_mpi0.5_L32}
\end{figure}

Figure~\ref{fig:our_method_log_sys_mpi0.5_L32} shows the $g_2^l$ dependence of the results obtained with the logarithmic function for $G(Q^2)$.
The result decreases with decreasing $g_2^l$ and becomes stable for $g_2^l \le 1$, which is the same behavior observed in the mock-data analysis in Sec.~\ref{sec:method:our_method}.
Thus, we regard the result with $g_2^l = 1$ as our central analysis, and the systematic error is estimated from the maximum difference between the central values and the others in the stable region $g_2^l < 1$.

To estimate the systematic error associated with $g_1^l$ in the logarithmic choice for $G(Q^2)$, we follow essentially the same procedure as in the quadratic case, but replace $g_2$ by $g_1^l$ and employ $g_2^l = 1$.

\begin{figure}[ht!]
    \includegraphics*[scale=1.0]{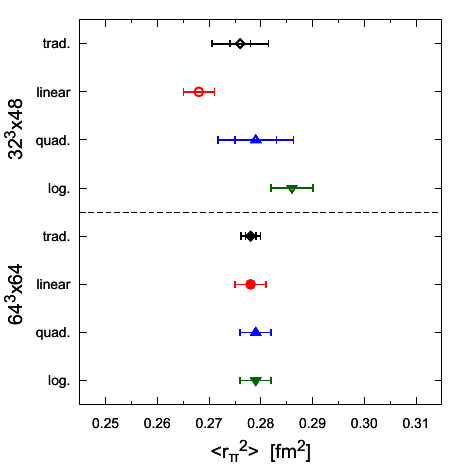}
    \caption{All the results of $\langle r^2_\pi \rangle$ at $m_\pi=0.5$ GeV. The labels ``trad.'', ``linear'', ``quad.'', and ``log.'' represent the results from the traditional, linear combination, and our improved methods with the quadratic and logarithmic functions, respectively.}
    \label{fig:comp_r2_mpi0.5}
\end{figure}

The results from our improved methods with the combined error with the statistical and systematic ones are plotted in Fig.~\ref{fig:comp_r2_mpi0.5}, together with that from the linear combination method.
Table~\ref{tab:r2_model-indep_mpi0.5_L32_L64} 
summarizes all the results on $L=32$ for the three methods: the linear combination method and our methods with quadratic and logarithmic functions.

\begin{table}[ht!]
    \caption{Results of the charge radius at $m_\pi=0.5$ GeV obtained using the model-independent methods. The labels for ``linear'', ``quad.'', and ``log.'' represent the linear combination and our improved methods with the quadratic and logarithmic functions, respectively. For our methods, the first error is statistical, while the other two are systematic errors for the choice of the parameter in $G(Q^2)$.}
    \label{tab:r2_model-indep_mpi0.5_L32_L64}
    \begin{tabular}{cccc}
        \hline
        \hline
         size  & linear & quad. & log. \\
         \hline
        $L=32$\,\, & $0.268(3)_{\text{stat.}}$ & $0.279(4)_{\text{stat.}}(6)_{g_{1}\text{sys.}}(1)_{g_{2}\text{sys.}}$ & $0.286(4)_{\text{stat.}}(1)_{g_{1}^{l}\text{sys.}}(0)_{g_{2}^{l}\text{sys.}}$ \\
        $L=64$\,\, & $0.278(3)_{\text{stat.}}$ & $0.279(3)_{\text{stat.}}(0)_{g_{1}\text{sys.}}(0)_{g_{2}\text{sys.}}$ & $0.279(3)_{\text{stat.}}(0)_{g_{1}^{l}\text{sys.}}(0)_{g_{2}^{l}\text{sys.}}$ \\
        \hline
        \hline
    \end{tabular}
\end{table}

We perform the same analyses with the model-independent methods using the data of the moments of the 3-point function on $L=64$.
The results with the three methods on $L=64$ are tabulated in Table~\ref{tab:r2_model-indep_mpi0.5_L32_L64} as well as those on $L=32$.

All the results of $\langle r^2_\pi \rangle$ explained above are compared in Fig.~\ref{fig:comp_r2_mpi0.5} together with those from the traditional method.
All the results on $L=64$ are in good agreement with each other.
Since the four results are stable against the calculation method and the physical spatial extent is $5.8$ fm on $L=64$, these results can be regarded as the results in the infinite volume.

For the traditional method on $L=32$, although the systematic error coming from the choice of the fit forms is much larger than that on $L=64$, the result agrees with the $L=64$ results within the large combined error.

The result from the linear combination method on $L=32$ is smaller by about 4\% than the $L=64$ results, while its statistical error is smaller than the combined error of the traditional method on $L=32$.
This discrepancy from the $L=64$ results is interpreted to be caused by the higher-order contribution in the linear combination method, because the size of the discrepancy roughly agrees with the one observed in the mock-data analysis with a similar parameter set in Sec.~\ref{sec:mockup_improved_method}.

For our improved method with the quadratic function, the central value agrees with the $L=64$ results, albeit its combined error is as large as that of the traditional method on $L=32$.
It suggests that the systematic error coming from the higher-order contribution becomes smaller than that in the linear combination method.
On the other hand, a non-negligible systematic error arises from the choice of the parameters of $G(Q^2)$, especially for $g_1$.
This trend is not observed in the mock-data analysis in Sec.~\ref{sec:mockup_improved_method}.
Thus, we expect that this is caused by effects not contained in the mock data generation, such as the correction of $F_\pi(Q^2)$ from the monopole form.
This possibility is investigated in Appendix~\ref{app:sys_err_mockup} using modified mock data to match the $F_\pi(Q^2)$ data shown in Fig.~\ref{fig:ff_pi_mpi0.5_L32}; the size of the $g_1$ dependence of about 2\% also appears in the modified mock-data analysis.
This systematic error of the $g_1$ dependence is sufficiently suppressed in the calculation on $L=64$ as shown in Fig.~\ref{fig:comp_r2_mpi0.5} and also Table~\ref{tab:r2_model-indep_mpi0.5_L32_L64}.

Another improved method with the logarithmic function gives a slight overestimation from the $L=64$ results, while the systematic error from the choice of the parameter in $G(Q^2)$ is much smaller than that in the quadratic function case.
Thus, the overestimation seems to be caused by the higher-order contribution as in the linear combination method.
The size of the systematic error, however, is smaller compared to that of the linear combination method on $L=32$.
Again, although this tendency is not observed in the mock-data analysis in Sec.~\ref{sec:mockup_improved_method}, a more realistic mock-data analysis in Appendix~\ref{app:sys_err_mockup} reasonably reproduces the overestimation of about 2\% from the $L=64$ results.
This systematic error becomes negligible on the larger volume, as in the linear combination method.

\subsection{Results at $m_\pi=0.3$ GeV}

\begin{figure}[ht!]
    \includegraphics*[scale=1.0]{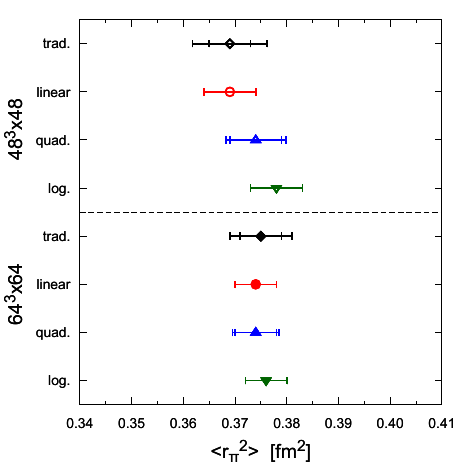}
    \caption{The same format as Fig.~\ref{fig:comp_r2_mpi0.5}, but for $m_\pi=0.3$ GeV.}
    \label{fig:comp_r2_mpi0.3}
\end{figure}

The same calculations as the $m_\pi=0.5$ GeV case are performed at $m_\pi=0.3$ GeV on $L=48$ and $64$ at the same lattice spacing.
The details of the calculation are described in Appendix~\ref{app:result_mpi0.3}.
A comparison of the results obtained from the four methods on the two volumes is shown in Fig.~\ref{fig:comp_r2_mpi0.3}.
In this $m_\pi$ case, all the results are in good agreement within their errors, including the linear combination method.
The figure suggests that our method works also at smaller $m_\pi$, and the finite-volume effect in the linear combination method is suppressed enough on $L=48$ in this $m_\pi$ case.

\section{Summary and discussion}
\label{Summary}

In this study, we have discussed an improved model-independent strategy to determine meson charge radii from correlation functions.
The model-independent method based on spatial moments does not require the assumption of a specific form of the form factor.
However, the correspondence between moments and derivatives is subject to non-negligible contamination in a finite spatial volume, as described in Ref.~\cite{Feng:2019geu}.
Specifically, residual contributions from higher-order terms in the Taylor expansion, which are the $n \ge 3$ terms in Eqs.~(\ref{eq:imp_method_R}) and (\ref{eq:our_method_R}), can introduce a systematic bias when the spatial extent is insufficiently large compared to the corresponding charge radius.

Our improvement of the model-independent method is conceptually straightforward.
We introduce an auxiliary function $G(Q^2)$ and redefine the function to be Taylor-expanded by considering the product $S(Q^2)=F(Q^2)G(Q^2)$ instead of $F(Q^2)$ itself.
The first Taylor coefficient of the form factor is then obtained through the same type of linear combination as in the construction of Feng {\it et al.}~\cite{Feng:2019geu}, with an explicit subtraction of the $g_1$ contribution arising from the expansion of $G(Q^2)$.

We have investigated two convenient parameterizations of $G(Q^2)$: a quadratic form, $G(Q^2)=1+g_1 Q^2+g_2 Q^4$, and a logarithmic form, $G(Q^2)=1+g_1^{l} \log(1+g_2^{l} Q^2)$.
The practical criterion for selecting the parameters in $G(Q^2)$ is to suppress the leading higher-order contamination by enforcing $s_2 \simeq 0$.
In practice, this condition is implemented by requiring that $R_2(t)=0$, as defined in Eq.~(\ref{eq:def_R2t}).
Both parameterizations were tested using mock data as well as actual lattice QCD data.
In both cases, we found that they effectively suppress higher-order contamination and provide reliable determinations of the target Taylor coefficient.

In practical applications, an important question is which choice of $G(Q^2)$ should be adopted as a default.
From our tests, both the quadratic and logarithmic parameterizations provide consistent determinations of the target Taylor coefficient on sufficiently large volumes, while their behaviors can differ mildly on smaller volumes.
In particular, for the $m_\pi=0.5$ GeV data on $L=32$, the logarithmic choice tends to yield a slightly larger central value than the values at large volume, whereas the quadratic choice yields a central value consistent with the large-volume value with a more conservative (larger) quoted uncertainty once the $g_1$ and $g_2$ systematics are included.
This suggests that the quadratic choice can be advantageous when one aims at a conservative error estimate, while the logarithmic choice can be advantageous when one prioritizes numerical stability and robustness of parameter determination.
Based on these considerations, we recommend the quadratic parameterization as the primary choice, while it is necessary to carefully choose $g_2$ from $R_2(t) = 0$ by avoiding the divergence of $1/G(Q^2)$. Another parameterization based on the logarithmic function can be useful for a cross-check of the result.

For future directions, it is possible to discuss further improvements of the model-independent method.
For example, in our strategy, one of the parameters in $G(Q^2)$ is fixed by hand, while it can be determined from the data by adding another condition.
Moreover, important future works are applications of those model-independent methods to charge radius calculations at the physical $m_\pi$ and also to determining the radius for various nucleon form factors.

\section*{Acknowledgments}
Numerical calculations in this work were performed on Oakforest-PACS and Wisteria/BDEC01(Odyssey) in Joint Center for Advanced High Performance Computing (JCAHPC) under Multidisciplinary Cooperative Research Program of Center for Computational Sciences, University of Tsukuba.
This research also used computational resources of Oakforest-PACS by Information Technology Center of the University of Tokyo, and of Fugaku by RIKEN CCS through the HPCI System Research Project (Project ID: hp170022, hp180051, hp180072, hp180126, hp190025, hp190081, hp200062, hp200167, hp210112, hp220079, hp230199, hp240207, hp250218).
The calculation employed OpenQCD system\footnote{http://luscher.web.cern.ch/luscher/openQCD/}.
This work was supported in part by Grants-in-Aid for Scientific Research from the Ministry of Education, Culture, Sports, Science and Technology (Nos. 19H01892, 23H01195, 23K25891), and
MEXT as ``Program for Promoting Researches on the Supercomputer Fugaku'' (Search for physics beyond the standard model using large-scale lattice QCD simulation and development of AI technology toward next-generation lattice QCD; Grant Number JPMXP1020230409), and JST SPRING, Japan Grant Number JPMJSP2124, and JST, the establishment of university fellowships towards the creation of science technology innovation, Grant Number JPMJFS2106.
The work of H.W. was partially supported by JSPS KAKENHI Grant Number 21J13014, 23K22489, and 24K00630.
This work is supported by the JLDG constructed over the SINET6 of NII.

\bibliographystyle{apsrev4-1}
\bibliography{reference}

\appendix

\section{Size of finite-volume effect in model-independent methods}
\label{app:sys_err_mockup}

In this appendix, the finite-volume effect in model-independent methods is estimated with mock data discussed in Sec.~\ref{sec:mockup}.
In addition to the three methods discussed in the section, the naive method in Eq.~(\ref{eq:naive_mod-indep_fv}) is also discussed in this appendix.
In the naive method, the parameter $A$ in Eq.~(\ref{eq:mockup_form_factor}) is evaluated by
\begin{equation}
    A_{\rm naive} = \frac{1}{2} D^{(1)}(t) - \frac{1}{2m_\pi}
    \left( \frac{1}{2m_\pi} + t \right) .
    \label{app:eq:naive}
\end{equation}
The size of the finite-volume effect is investigated by
a relative difference of $A$ defined as
\begin{equation}
    \delta_J = \left| 1 - \frac{A_J}{A}\right|,
\end{equation}
where $A_J$ is a result obtained from a model-independent method for $J =$ naive, lin, quad, and log, corresponding to the naive, linear combination, and our methods with the quadratic and logarithmic functions.
The parameters in our methods are determined in the same way as explained in Sec.~\ref{sec:mockup}.

\subsection{$m_\pi$ dependence}

An $m_\pi$ dependence of $\delta_J$ is investigated using the mock data that are generated with $A = 5$ and $t = 25$.
We employ $m_\pi = 0.25$ and 0.15 corresponding to $m_\pi=0.5$ GeV and 0.3 GeV in the actual lattice QCD data in Sec.~\ref{sec:Numerical validation with actual lattice QCD configurations}.

\begin{figure}[ht!]
    \includegraphics*[scale=.45]{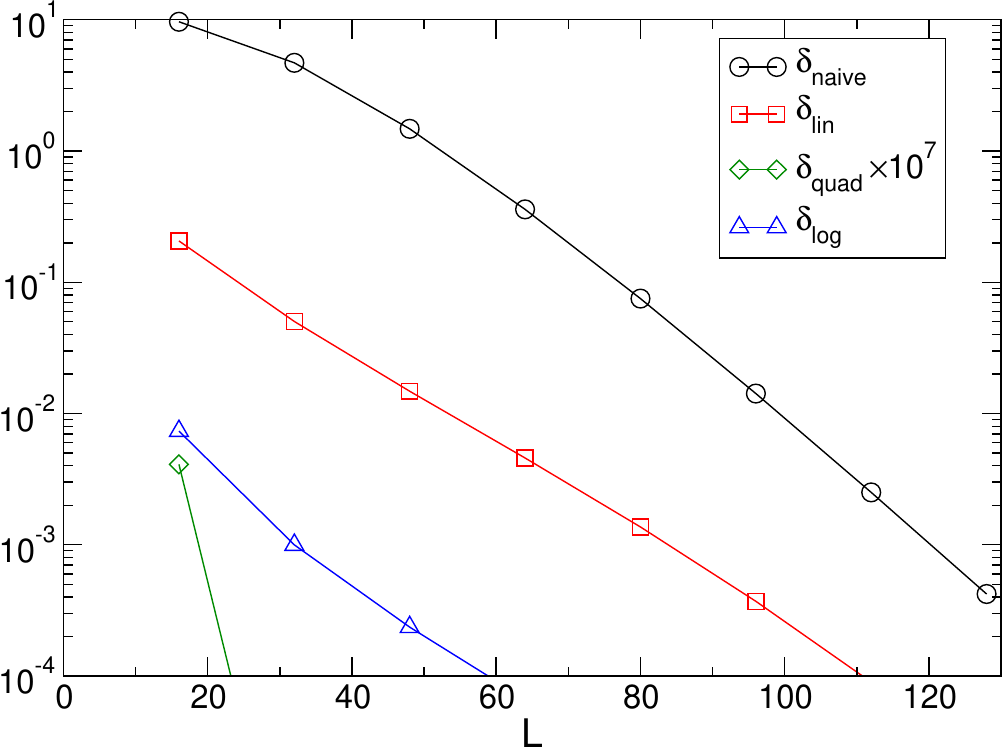}
    \includegraphics*[scale=.45]{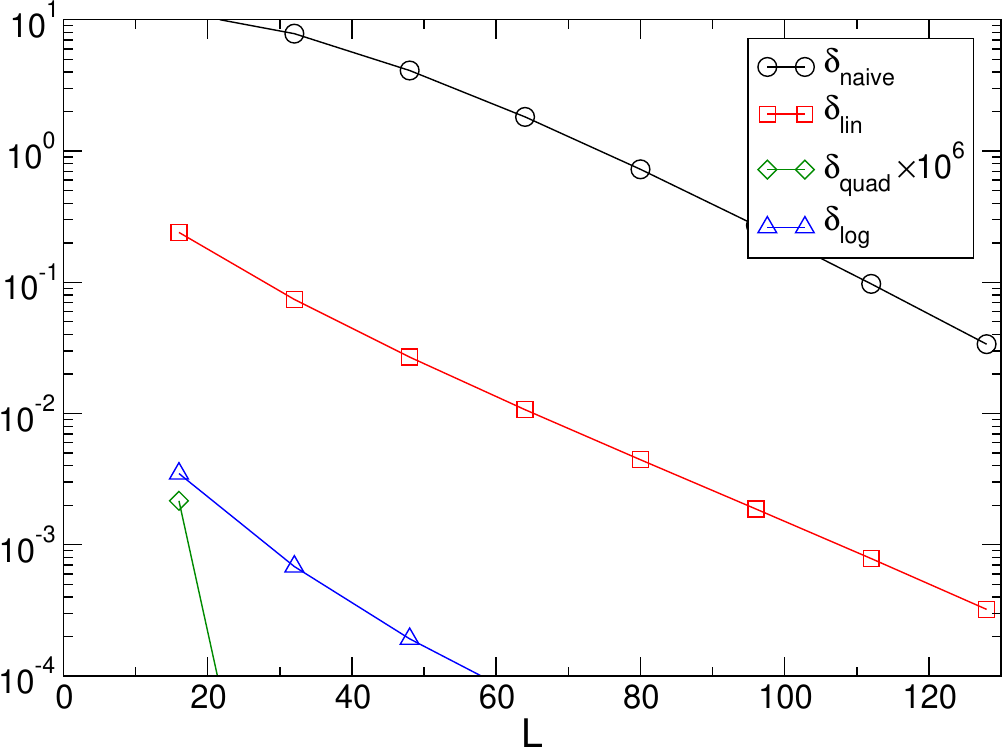}
    \caption{Relative difference $\delta_J$ for the mock data with $A=5$ and $t=25$ as a function of $L$.
    The left and right panels correspond to $m_\pi = 0.25$ and 0.15, respectively.}
    \label{fig:appA:comp_m0.5+m0.3}
\end{figure}

Figure~\ref{fig:appA:comp_m0.5+m0.3} presents $\delta_J$ as a function of $L$ for the four different methods.
As expected, all $\delta_J$ in both $m_\pi$ cases decrease as $L$ increases.
The finite-volume effect is largely reduced in the linear combination method compared to the naive one.
It is further improved in our methods, especially using the quadratic
function.

The right panel of Fig.~\ref{fig:appA:comp_m0.5+m0.3} shows that $\delta_J$ in $m_\pi = 0.15$ are larger than those in $m_\pi = 0.25$, except for $\delta_{\rm log}$, which is similar to the one in $m_\pi = 0.25$.

\subsection{Modified monopole form factor for $L=32$ and $m_\pi=0.5$ GeV}

Our improved method with the logarithmic function gives a larger $\langle r^2 \rangle$ on $L=32$ at $m_\pi=0.5$ GeV than that on $L=64$, as shown in Fig.~\ref{fig:comp_r2_mpi0.5}.
To understand this systematic error, we employ a modified monopole form for $F_\pi(Q^2)$ given by
\begin{equation}
    F_\pi(Q^2) = \frac{ 1 + C Q^6 }{ 1 + AQ^2 }, 
    \label{eq:appA:mod_monopole}
\end{equation}
where we choose $A = 5.75$ and $C = 12$ to reproduce the $F_\pi(Q^2)$ data on $L=32$ presented in Fig.~\ref{fig:ff_pi_mpi0.5_L32}, especially in the larger $Q^2$ region.
The parameters, $m_\pi = 0.25$ and $t=10$, are selected to match the $L=32$ data.

\begin{figure}[ht!]
    \includegraphics*[scale=.45]{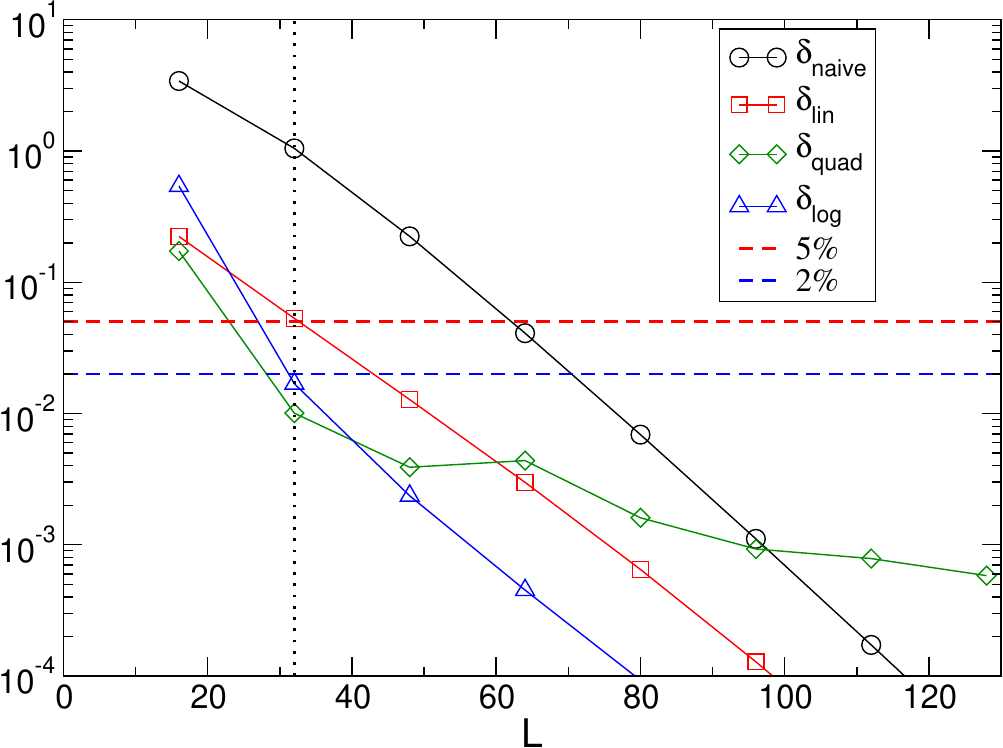}
    \caption{Results of $\delta_J$ with the mock data with the modified monopole form factor in Eq.~(\ref{eq:appA:mod_monopole}).
    The red and blue lines express 5\% and 2\% relative differences on $L = 32$.}
    \label{fig:appA:comp_m0.5_mod}
\end{figure}

The mock data of the moment of the correlation function is generated, including the effect of $R_Z(Q^2) \ne 1$ as in Eq.~(\ref{eq:1d-3pt_Dk}).
In this test, we choose $R_Z(Q^2) = 1 + Z_1 Q^2 + Z_2 Q^4$ with $Z_1 = 0.1$ and $Z_2 = 1.5$ to reproduce the $R_Z(Q^2)$ data in Fig.~\ref{fig:model-indep_Zpi_ov_Zp_mpi0.5_L32}.
To remove the $Z_1$ effect from the evaluated value of $A_J$, we replace $A_J$ by $A_J + Z_1$ for all the cases.

The results of $A_J$ with the different model-independent methods are listed in Table~\ref{tab:appA:mod_monopole:r2_model_indep}. 
It suggests that our method with the logarithmic function gives a 1.7\% larger value than the input value of $A$, whose size is comparable with the result of the lattice QCD data analysis on $L=32$ discussed in Sec.~\ref{sec:result:ana_L32_m0.5:r2}.
Our method with the quadratic function and the linear combination method have negative systematic errors of 1.0\% and 5.4\%, respectively.
Those values are also compatible with the results of the lattice QCD data.
The table also shows that the size of the systematic error decreases as the volume increases, as expected. 
The systematic errors in the three methods become less than 0.5\% on $L=64$.

The results of $\delta_J$ are plotted in Fig.~\ref{fig:appA:comp_m0.5_mod}, including that from the naive method.
The linear combination method gives values of $\delta_J$ similar to those of the monopole form case as in Fig.~\ref{fig:appA:comp_m0.5+m0.3} with the same $m_\pi$.
In contrast to this similarity, $\delta_{\rm quad}$ and $\delta_{\rm log}$ are highly sensitive to a correction of the monopole form, because they differ by several orders of magnitude from those in Fig.~\ref{fig:appA:comp_m0.5+m0.3}.
Especially, $\delta_{\rm quad}$ shows poor convergence with $L$, although it is below 1\% in the large-$L$ region.

We also investigate the $g_1$ dependence of the result from our method with the quadratic function.
Table~\ref{tab:appA:mod_monopole:r2_model_indep} summarizes the values using the analysis with $g_1 = A$ and $-A$ instead of our standard choice of $g_1 = 0$.
The 2.5\% difference between the results with $g_1 = -A$ and $g_1 = 0$ (denoted by quad.) on $L=32$ is reasonably consistent with that observed in the lattice QCD analysis on $L=32$ as shown in Fig.~\ref{fig:our_method_quad_sys_mpi0.5_L32}.

\begin{table}[ht!]
    \caption{Results of $A_J$ obtained from the model-independent methods using the modified monopole form factor data with the input $A=5.75$. The labels for ``linear'', ``quad.'', and ``log.'' represent the linear combination and our improved methods with the quadratic and logarithmic functions, respectively.
    The results obtained from our method with the quadratic function using $g_1 = \pm A$ are also tabulated.}
    \label{tab:appA:mod_monopole:r2_model_indep}
    \begin{tabular}{cccccc}
        \hline
        \hline
        size  & linear & quad. & log. & $g_1=A$ & $g_1=-A$ \\
        \hline
        $L=32$\,\, & 5.44 & 5.69 & 5.85 & 5.62 & 5.83 \\
        $L=64$\,\, & 5.73 & 5.72 & 5.75 & 5.75 & 5.73 \\
        \hline
        \hline
    \end{tabular}
\end{table}

\subsection{Parameters for the physical $m_\pi$}

The results of $\delta_J$ are evaluated using the mock data with the parameters for the physical point, $m_\pi = 0.135$ GeV and $\langle r^2 \rangle = 0.44$ fm$^2$ at $a = 0.08$ fm.
We set the reference time $t = 20$, which is a similar value to that in Ref.~\cite{Sato:2024epw}.
In this calculation, the exact monopole form and $R_Z(Q^2) = 1$ are adopted.

\begin{figure}[ht!]
    \includegraphics*[scale=.45]{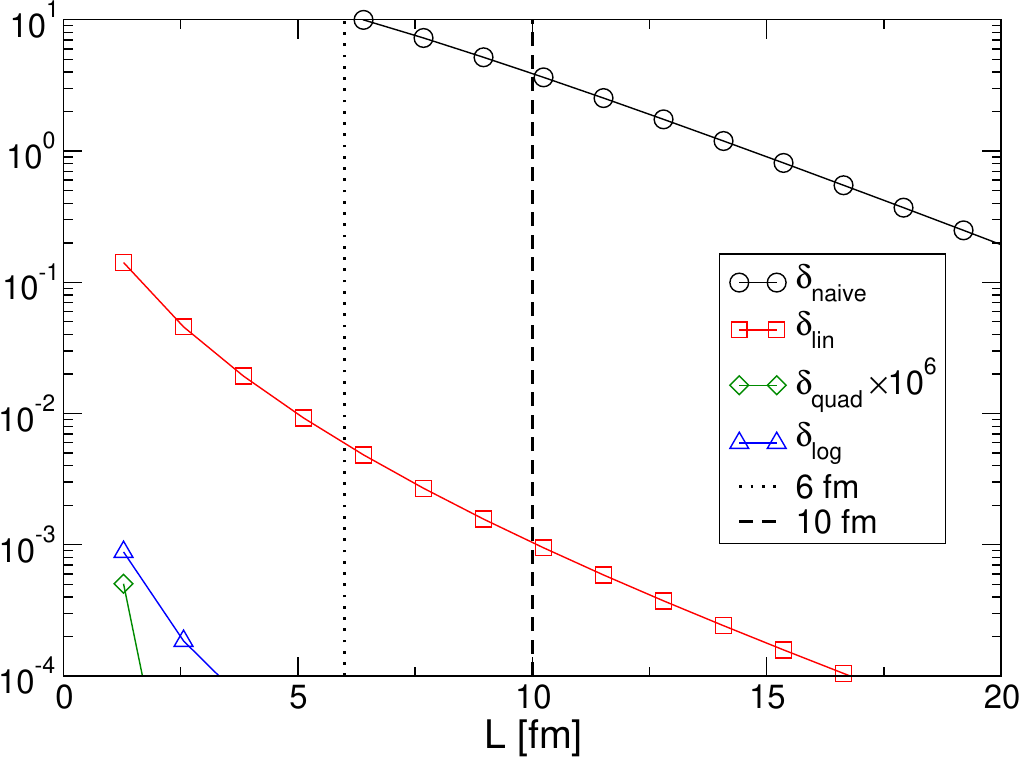}
    \caption{Results of $\delta_J$ with the mock data with the parameters for the physical $m_\pi$ as a function of the physical spatial extent $L$ at $a = 0.08$ fm.
    The dot and dashed vertical lines represent $L = 6$ and $10$ fm, respectively.}
    \label{fig:appA:comp_pacs10}
\end{figure}

Figure~\ref{fig:appA:comp_pacs10} shows that $\delta_{\rm lin}$ is roughly 1\% and 0.1\% at $L = 6$ and 10 fm, respectively, corresponding to the volumes for the calculations in Refs.~\cite{Feng:2019geu} and \cite{Sato:2024epw}.
The results for $\delta_{\rm quad}$ and $\delta_{\rm log}$ are significantly smaller than $\delta_{\rm lin}$.
However, it is expected that they would change significantly if the form factor has a correction from the monopole as discussed in the previous subsection.

\section{Results at $m_\pi=0.3$ GeV}
\label{app:result_mpi0.3}

We also perform the same analysis on the $m_\pi=0.3$ GeV ensembles.
These gauge configurations were generated with the same setup as in the $m_\pi=0.5$ GeV case, namely $N_f=2+1$ dynamical flavors, the Iwasaki gauge action, and the nonperturbative $O(a)$-improved Wilson quark action.
The lattice sizes are $L^3\times T = 48^3\times 48$ and $64^3\times 64$.
The details of the gauge ensembles are described in Ref.~\cite{Yamazaki:2015asa}.
The ensemble parameters are summarized in Table~\ref{tab:sim_para_mpi0.3}, and the number of measurements per configuration is listed in Table~\ref{tab:meas_para_mpi0.3}.

\begin{figure}[ht!]
    \includegraphics*[scale=1.0]{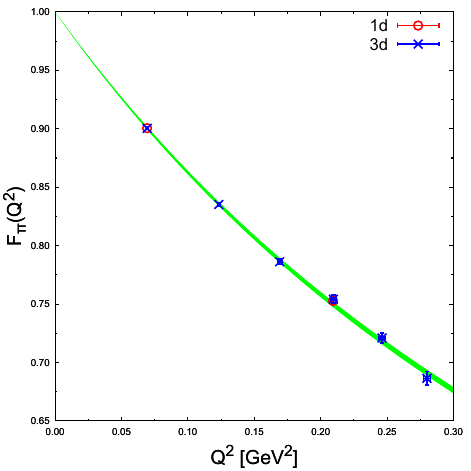}
    \caption{Result of $F_\pi(Q^2)$ at $m_\pi=0.3$ GeV on $L = 48$ as a function of $Q^2$.}
    \label{fig:ff_pi_mpi0.3_L48}
\end{figure}

\begin{figure}[ht!]
    \includegraphics*[scale=1.0]{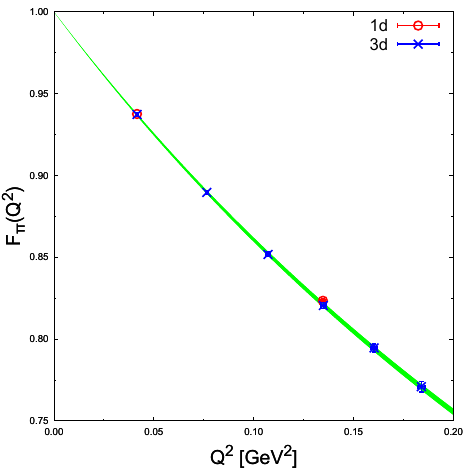}
    \caption{Result of $F_\pi(Q^2)$ at $m_\pi=0.3$ GeV on $L = 64$ as a function of $Q^2$.}
    \label{fig:ff_pi_mpi0.3_L64}
\end{figure}

\begin{figure}[ht!]
    \includegraphics*[scale=1.0]{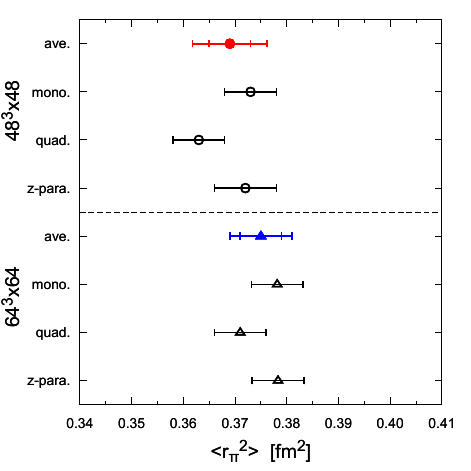}
    \caption{Results of $\langle r^2_\pi \rangle$ obtained from a fit to $F_\pi(Q^2)$ with each fit form and their weighted average at $m_\pi=0.3$ GeV.}
    \label{fig:trad_r2_mpi0.3}
\end{figure}

The measured $F_\pi(Q^2)$ on $L=48$ and $L=64$ are shown as functions of $Q^2$ in Figs.~\ref{fig:ff_pi_mpi0.3_L48} and \ref{fig:ff_pi_mpi0.3_L64}, respectively.
The numerical values of $F_\pi(Q^2)$ are summarized in Tables~\ref{tab:ff_mpi0.3_L48} and \ref{tab:ff_mpi0.3_L64}.
The $F_\pi(Q^2)$ data are analyzed using three fit forms: a monopole form, a quadratic polynomial in $Q^2$, and the $z$-parameter expansion,
\begin{equation}
    F_\pi(Q^2) = \left\{
    \begin{array}{l}
        \left(1+AQ^2\right)^{-1},\\
        1 - A Q^2 + B Q^4,\\
        1 - a_1 z + a_2 z^2,
    \end{array}
    \right.
\end{equation}
where $A$, $B$ and $a_1$, $a_2$ are fit parameters.
The definition of the $z$ variable in the $z$-parameter expansion is the same as that given in Eq.~(\ref{eq:z-para}).
The fit results obtained with the monopole form are also shown in Figs.~\ref{fig:ff_pi_mpi0.3_L48} and \ref{fig:ff_pi_mpi0.3_L64}.
The resulting values of $\langle r_\pi^2\rangle$ on $L=48$ and $L=64$ are summarized in Tables~\ref{tab:trad_r2_fit_results_mpi0.3_L48} and \ref{tab:trad_r2_fit_results_mpi0.3_L64}, respectively.
For the fits on $L=48$, all six $Q^2$ data points are used in the monopole fit, whereas the quadratic polynomial fit and the $z$-expansion fit are performed using only the three lowest-$Q^2$ points.
On the other hand, for $L=64$, all available $Q^2$ data points are used in all fit forms.
The results of $\langle r^2_\pi \rangle$ are summarized in Fig.~\ref{fig:trad_r2_mpi0.3}.

The results obtained with the model-independent methods are also summarized in Table~\ref{tab:r2_model-indep_mpi0.3_L48_L64}.
The analysis procedure is the same as that employed in the $m_\pi=0.5$~GeV case.

\begin{table}[t!]
    \caption{Simulation parameters for $m_\pi=0.3$ GeV ensembles.}
    \label{tab:sim_para_mpi0.3}
    \begin{tabular}{cccccccc}
        \hline
        \hline
        $\beta$ & $L^3\times T$ & $a^{-1}$[GeV] & $a$[fm] & $L$[fm] & $m_\pi$[GeV] & $N_{\rm conf}$ \\
        \hline
        1.90 & 48$^3\times$48 & 2.194(10) & 0.08995(40) & 4.3 & 0.301(2)  & 100 \\
        1.90 & 64$^3\times$64 & 2.194(10) & 0.08995(40) & 5.8 & 0.302(2)  & 80 \\
        \hline
        \hline
    \end{tabular}
\end{table}

\begin{table}[t!]
    \caption{Details of the measurements for $m_\pi=0.3$ GeV ensembles.
    $N_{\rm meas}$ is counted in the same way as when $m_\pi=0.5$ GeV in Table~\ref{tab:meas_para}.}
    \label{tab:meas_para_mpi0.3}
    \begin{tabular}{cccccccc}
        \hline
        \hline
        Type of data & $L^3\times T$ & $N_{\rm src}$ & $N_{\rm r}$ & $N_{\rm meas}$ \\
        \hline
        1d & 48$^3\times$48 & 32 & 6 & 2304 \\
           & 64$^3\times$64 & 32 & 4 & 1536 \\
        \hline
        3d & 48$^3\times$48 & 4 & 27 & $432 \times N_{d}$ \\
           & 64$^3\times$64 & 4 & 10 & $160 \times N_{d}$ \\
        \hline
        \hline
    \end{tabular}
\end{table}

\begin{table}[t!]
    \caption{Result of $F_\pi(Q^2)$ for each $Q^2$ at $m_\pi=0.3$ GeV on $L = 48$.}
    \label{tab:ff_mpi0.3_L48}
    \begin{tabular}{cc}
        \hline
        \hline
        $Q^2$[GeV$^2$] & $F_\pi(Q^2)$ \\
        \hline
        0.0693(06) & 0.900(1) \\
        0.1233(11) & 0.835(1) \\
        0.1691(16) & 0.786(2) \\
        0.2095(20) & 0.754(3) \\
        0.2462(23) & 0.721(4) \\
        0.2799(26)  & 0.686(6) \\
        \hline
        \hline
    \end{tabular}
\end{table}

\begin{table}[t!]
    \caption{Result of $F_\pi(Q^2)$ for each $Q^2$ at $m_\pi=0.3$ GeV on $L = 64$.}
    \label{tab:ff_mpi0.3_L64}
    \begin{tabular}{cc}
        \hline
        \hline
        $Q^2$[GeV$^2$] & $F_\pi(Q^2)$ \\
        \hline
        0.0416(04) & 0.9373(06) \\
        0.0765(07) & 0.8897(09) \\
        0.1071(10) & 0.8518(14) \\
        0.1348(12) & 0.8205(19) \\
        0.1602(15) & 0.7946(25) \\
        0.1839(17) & 0.7709(33) \\
        \hline
        \hline
    \end{tabular}
\end{table}

\begin{table}[t!]
    \caption{Results of $\langle r^2_\pi \rangle$ obtained from a fit to $F_\pi(Q^2)$ with each fit form and their weighted average at $m_\pi=0.3$ GeV on $L = 48$.}
    \label{tab:trad_r2_fit_results_mpi0.3_L48}
    \begin{tabular}{ccc}
        \hline
        \hline
        fit form  & $\langle r^2_\pi \rangle$[fm$^2$] & $\chi^2$/d.o.f. \\
        \hline
        monopole  & 0.373(5) & 0.7 \\
        quadratic & 0.363(5) & 0.2 \\
        $z$-exp   & 0.372(6) & 0.007 \\
        average   & $0.369(4)_{\rm stat.}(6)_{\rm sys.}$ & - \\
        \hline
        \hline
    \end{tabular}
\end{table}

\begin{table}[t!]
    \caption{Results of $\langle r^2_\pi \rangle$ obtained from a fit to $F_\pi(Q^2)$ with each fit form and their weighted average at $m_\pi=0.3$ GeV on $L = 64$.}
    \label{tab:trad_r2_fit_results_mpi0.3_L64}
    \begin{tabular}{ccc}
        \hline
        \hline
        fit form  & $\langle r^2_\pi \rangle$[fm$^2$] & $\chi^2$/d.o.f. \\
        \hline
        monopole  & 0.378(5) & 0.1 \\
        quadratic & 0.371(5) & 0.2 \\
        $z$-exp   & 0.378(5) & 0.7 \\
        average   & $0.375(4)_{\rm stat.}(5)_{\rm sys.}$ & - \\
        \hline
        \hline
    \end{tabular}
\end{table}

\begin{table}[ht!]
    \caption{Results of the charge radius at $m_\pi=0.3$ GeV obtained using the model-independent methods. The format is the same as in Table~\ref{tab:r2_model-indep_mpi0.5_L32_L64}.}
    \label{tab:r2_model-indep_mpi0.3_L48_L64}
    \begin{tabular}{cccc}
        \hline
        \hline
         size  & linear & quad. & log. \\
        \hline
        $L=48$\,\, & $0.369(5)_{\text{stat.}}$ & $0.374(5)_{\text{stat.}}(3)_{g_{1}\text{sys.}}(0)_{g_{2}\text{sys.}}$ & $0.378(5)_{\text{stat.}}(0)_{g_{1}^{l}\text{sys.}}(0)_{g_{2}^{l}\text{sys.}}$ \\
        $L=64$\,\, & $0.373(4)_{\text{stat.}}$ & $0.374(4)_{\text{stat.}}(2)_{g_{1}\text{sys.}}(0)_{g_{2}\text{sys.}}$ & $0.376(4)_{\text{stat.}}(0)_{g_{1}^{l}\text{sys.}}(0)_{g_{2}^{l}\text{sys.}}$ \\
        \hline
        \hline
    \end{tabular}
\end{table}

\end{document}